\documentclass[lettersize,journal]{IEEEtran}
\usepackage{amsmath,amsfonts}
\usepackage{amsmath}
\usepackage{algpseudocode}
\usepackage{algorithm}
\usepackage{algpseudocode}
\usepackage{array}
\usepackage[caption=false,font=normalsize]{subfig}
\usepackage{textcomp}
\usepackage{stfloats}
\usepackage{url}
\usepackage{verbatim}
\usepackage{gensymb}
\usepackage{graphicx}
\usepackage{cite}
\usepackage{siunitx}
\usepackage{tabularx}
\usepackage{multicol}
\usepackage[export]{adjustbox}
\usepackage{threeparttable}

\usepackage{pifont}

\usepackage[table,xcdraw]{xcolor}
\usepackage{multirow}
\usepackage{ifthen}
\usepackage{tikz}
\usepackage{tikzsymbols}
\usetikzlibrary{positioning, shapes.geometric, shapes.symbols, arrows, automata, spy}
\usepackage{pgfplots,pgfplotstable} 

\usepackage{wasysym}
\usepackage{scalerel}
\usepackage{tikz}
\usetikzlibrary{svg.path}

\usepackage{soul}

\definecolor{color1}{RGB}{223, 225, 255}
\definecolor{color2}{RGB}{0, 0, 160}
\definecolor{color3}{RGB}{51, 51, 51}
\definecolor{color4}{RGB}{255, 255, 255}
\definecolor{color5}{RGB}{124, 191, 51}
\definecolor{color6}{RGB}{255, 216, 176}
\definecolor{color7}{RGB}{0, 0, 0}
\definecolor{color8}{RGB}{254, 226, 235}
\definecolor{color9}{RGB}{255, 138, 138}
\definecolor{color10}{RGB}{179, 254, 174}
\definecolor{color11}{RGB}{239, 239, 248}
\definecolor{color12}{RGB}{255, 192, 0}
\definecolor{color13}{RGB}{255, 255, 255}
\definecolor{color14}{RGB}{255, 179, 178}
\definecolor{color15}{RGB}{162, 177, 195}
\definecolor{color16}{RGB}{138, 138, 138}

\definecolor{color17}{RGB}{242, 248, 253}
\definecolor{color18}{RGB}{196, 199, 85}
\definecolor{color19}{RGB}{51, 51, 51}

\algnewcommand\algorithmicforeach{\textbf{foreach}}
\algdef{S}[FOR]{ForEach}[1]{\algorithmicforeach\ #1\ \algorithmicdo}

\usepackage[hidelinks,colorlinks=true,linkcolor=blue,citecolor=blue]{hyperref}
\usepackage{orcidlink}

\begin{document}

\clearpage
\onecolumn 
\pagestyle{empty} 
{© 2024 IEEE.  Personal use of this material is permitted.  Permission from IEEE must be obtained for all other uses, in any current or future media, including reprinting/republishing this material for advertising or promotional purposes, creating new collective works, for resale or redistribution to servers or lists, or reuse of any copyrighted component of this work in other works.
DOI: \href{https://doi.org/10.1109/TC.2024.3449082}{10.1109/TC.2024.3449082}

\twocolumn

\title{SCARF: Securing Chips with a Robust Framework against Fabrication-time Hardware Trojans}

\author{Mohammad Eslami\orcidlink{0000-0001-7200-3655}, ~\IEEEmembership{Graduate Student Member,~IEEE,}
Tara Ghasempouri\orcidlink{0000-0001-8021-9368}, ~\IEEEmembership{Member,~IEEE, \\}
Samuel Pagliarini\orcidlink{0000-0002-5294-0606}, ~\IEEEmembership{Member,~IEEE}
\thanks{This work was partially supported by the EU through the European Social Fund in the context of the project “ICT programme”. M. Eslami was also supported by HARNO (Grant No. 11.4-1/23/1).}
\thanks{M. Eslami, T. Ghasempouri, and S. Pagliarini are with the Department of Computer Systems, Tallinn University of Technology (TalTech), 12618, Tallinn, Estonia (e-mail: mohammad.eslami@taltech.ee; tara.ghasempouri@taltech.ee; samuel.pagliarini@taltech.ee)}
\thanks{Samuel Pagliarini is also with the ECE Department at Carnegie Mellon University, Pittsburgh, PA (e-mail: pagliarini@cmu.edu)}}


\maketitle

\begin{abstract}
The globalization of the semiconductor industry has introduced security challenges to Integrated Circuits (ICs), particularly those related to the threat of Hardware Trojans (HTs) -- malicious logic that can be introduced during IC fabrication. While significant efforts are directed towards verifying the correctness and reliability of ICs, their security is often overlooked. In this paper, we propose a comprehensive framework that integrates a suite of methodologies for both front-end and back-end stages of design, aimed at enhancing the security of ICs. Initially, we outline a systematic methodology to transform existing verification assets into potent security checkers by repurposing verification assertions. To further improve security, we introduce an innovative methodology for integrating online monitors during physical synthesis -- a back-end insertion providing an additional layer of defense. Experimental results demonstrate a significant increase in security, measured by our introduced metric, Security Coverage (SC), with a marginal rise in area and power consumption, typically under 20\%. The insertion of online monitors during physical synthesis enhances security metrics by up to 33.5\%. This holistic framework offers a comprehensive defense mechanism across the entire spectrum of IC design.
\end{abstract}

\begin{IEEEkeywords}
IC Design, ASIC, Hardware Trojan Horse, Verification, Assertions, Online Checkers, DfHT.
\end{IEEEkeywords}

\section{Introduction} \label{intro}
\IEEEPARstart{T}{he} fabrication of an Integrated Circuit (IC) is mostly performed in a fabless fashion, a model in which the fabrication of an IC is performed in other places rather than \emph{inside} the design house. Globalization has commonly led to the widespread adoption of this model by most companies, primarily driven by the impracticality of establishing proprietary fabrication facilities that require substantial financial investments in the order of billions of dollars. Notably, even industry leaders such as Apple opt to outsource the fabrication of their chips to external entities \cite{tsmc}. This strategic approach highlights the economic sensibility of relying on specialized foundries for chip production, especially for advanced node technologies, allowing design houses to channel resources more efficiently towards innovation and design endeavors.

Nevertheless, while this fabless model offers numerous advantages, it also comes with a tangible dark side. When the chip is sent to the foundry for fabrication, there will be no guarantee that the returned chip precisely aligns with the initial specifications (i.e., it can be modified inside the foundry). Even small modifications to the chip's design can compromise its security and pose potential risks to human life or lead to financial losses. This modification is commonly referred to as Hardware Trojan (HT), which is a malicious alteration, addition, or subtraction to the original design with the intent to compromise its integrity \cite{htsurv, tehrani10}. The goal of such manipulations can range from information leakage, functionality change, performance degradation, or the deliberate reduction of the chip's useful lifespan \cite{breakingSilic, chipREsurvey}. An HT is composed of two parts, namely, the trigger and the payload. Trigger, as its name indicates, is the activation mechanism of the HT, while the payload is the saboteur function that is executed when the trigger is activated. The trigger part can be designed to activate under specific temporal and/or temperature conditions or through a combination of certain inputs or internal signals of the chip. A sophisticated attacker strategically sets the trigger to activate only under extremely rare conditions, thus ensuring that the HT remains undetected by conventional detection schemes during normal chip operation \cite{htsurv, TrjClass}.

Therefore, design companies should proactively implement measures to safeguard their chips against fabrication-time inserted HTs \cite{HwSOverview, PrimeronHS, ProtectTrj, DETERRENT}. Moreover, the importance of this research cannot be emphasized enough. Historically, the primary focus during the verification stage within design houses has been on detecting and fixing bugs. Security, which is often overlooked in hardware development in favor of reliability and dependability, must no longer be neglected, given the surge in hardware attacks \cite{enisa}. 

Given this scenario, our focus is primarily on enhancing HT detection by incorporating defensive techniques at different stages of chip design, \textbf{both} in the front-end and back-end. This approach is rarely observed in prior works. In this paper, first, we show how the knowledge generated by verification engineers, specifically assertions, can be leveraged as valuable security assets during the front-end stage. These data are usually utilized to prove that a design is bug-free, and not used again. Repurposing this knowledge for security purposes has the potential to yield substantial time and effort savings.

However, relying solely on the reuse of verification data may not suffice to attain high levels of security. Therefore, we introduce a complementary technique aimed at enhancing the security of digital designs, which is the incorporation of online checkers during the back-end stage of physical synthesis. The insertion of the online monitors serves a dual purpose: not only do they aid in HT detection, but they also act as functional layout fillers. By utilizing the free resources within the layout, such as gaps and routing resources, these monitors contribute to restricting the available resources that an adversary might exploit for inserting HTs. Nonetheless, adding an online checker is a form of redundancy that may pose significant design overheads (in terms of timing, area, power, and performance). Yet, by proposing a smart timing- and area-aware technique, and performing the checker insertion by leveraging the Engineering Change Order (ECO) features of commercial CAD tools, we considerably minimize the overheads in our flow.

This smart technique seeks to make a balance between heightened security measures and the optimization of resource utilization, thereby enhancing the overall robustness of digital designs. Our overall methodology for reusing the verification data for security purposes, as the first contribution of this work, is described in our prior work~\cite{assertion}. In this paper, we present an in-depth description of how we make use of back-end stage techniques to augment our prior research on IC security.

The remaining sections of the paper are organized as follows: Section \ref{sec:bg} provides background information on various techniques against hardware attacks during fabrication time and reviews related works. Section \ref{sec:reuse} explores the primary motivation behind this work, focusing on transforming verification assertions into security checkers. In Section \ref{sec:online_monitors}, technical details regarding the implementation flow of adding online checkers during physical synthesis are explained. Experimental results are presented in Section \ref{sec:results}. The paper concludes with Section \ref{sec:conclusion}.

\begin{figure*}[t]
    \centering
    \includegraphics[width=1\textwidth, trim=.1cm .1cm .1cm .1cm, clip]{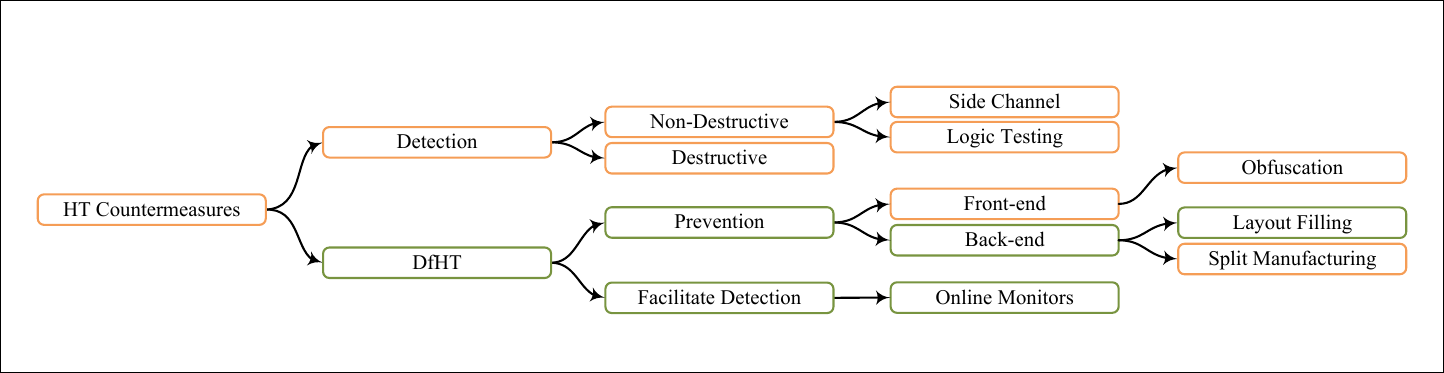}
     \vspace{-10mm}
    \caption{An overview of the HT protection methods. The techniques used in this work are colored in green.}
    \label{fig:contermeasures}
\end{figure*}

\section{Background} \label{sec:bg}
Due to the rising concerns regarding hardware attacks, design houses try to protect their chips, particularly during the post-design stages where the environment is not trusted, and there is limited oversight in the chip fabrication process. To address this particular oversight, programs such as the Microelectronics Quantifiable Assurance (MQA) have started to add traceability to devices and systems. In tandem, security concerns have also brought the concept of a ``Zero Trust'' model, in which all parties, tools, and assets thereof are considered potentially untrusted. These concepts co-exist with techniques to improve trust; An overview of the different protection techniques is shown in Fig.~\ref{fig:contermeasures}. These techniques are based on the concept of either detecting HTs on the fabricated chip or preventing the insertion of HTs through Design for Hardware Trust (DfHT) \cite{dfht} approaches. 

\subsection{Detection Techniques}
Once an adversary introduces an HT at fabrication time, detection becomes exceptionally challenging \cite{trojanLessons}. Effective detection techniques must be used after fabrication to counter this threat. These techniques are mainly based on inspecting the fabricated chip to ensure that the chip is HT-free. Detection methods are performed either in a destructive or non-destructive fashion. In destructive methods, the chip is de-packaged and each layer is separately inspected using highly advanced methods to check if any logic is added (or removed) by an adversary \cite{13}. 

However, this approach has significant drawbacks. Firstly, it involves significant time and cost. 
Secondly, the analyzed chip is rendered entirely non-operational after inspection, leading to its destruction. Consequently, this method is limited to analyzing random samples and is not viable for inspecting entire lots of chips.

Therefore, non-destructive methods were introduced to analyze the fabricated chips before being distributed in the market \cite{14,15,16,17,20,21,25}. These methods are mainly classified into side-channel analysis and logic testing. 

Side-channel analysis techniques concentrate on observing physical attributes such as power consumption, path delay, or electromagnetic emanations \cite{14,15,16,17}. These attributes are then compared to those of \emph{golden chips}, which are assumed to be HT-free. 

Logic testing involves applying test stimuli to evaluate chips, comparing the responses with expected ones precomputed through simulation \cite{20,21,25}. Detection relies on detecting changes in chip functionality during testing, and aims to activate potential HTs within a limited test time. To avoid detection, attackers create static HTs activated by extremely rare conditions. 

\subsection{DfHT Techniques}
In DfHT approaches, the concept is to embed additional protection logic in order to either aid in detecting the HTs \cite{46,49,52} or to prevent an adversary from inserting an HT altogether~\cite{35,33,37,bisa,40,papa,42}. Although complete prevention against HT insertion remains impossible in practice, efforts have focused on strategically limiting available chip resources, making it exceedingly difficult for adversaries to exploit them for malicious logic insertion \cite{bisa, 40, papa, 42}.

\begin{figure*}[!]
    \centering
    \includegraphics[width=.99\textwidth, trim=.1cm .1cm .1cm .1cm, clip]{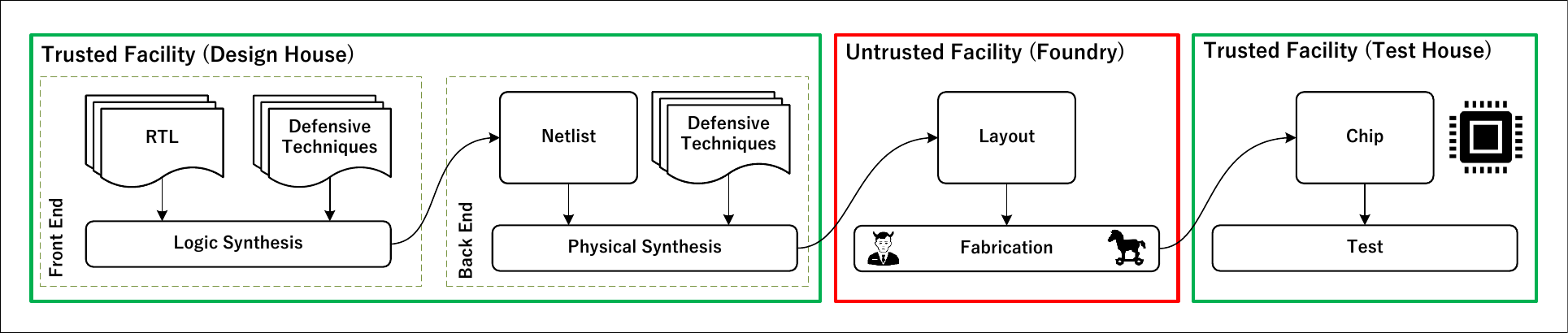}
     \vspace{-5mm}
    \caption{Different stages of IC design: The design house and the test house are considered trusted, while the foundry is assumed to be untrusted.}
    \label{fig:threat_model}
\end{figure*}

When referring to the process of designing an IC, the tasks are typically divided into two stages: front-end and back-end \cite{24a}. The front-end stage focuses on the initial specification and creation of the chip's architecture. This involves tasks such as specifying the functionality, creating a high-level design, and simulating the behavior of the design. Hardware Description Languages (HDLs) like Verilog or VHDL are commonly used in this stage. Once the front-end design is complete, the process transitions to the back-end. In this stage, the focus is on transforming the design into a physical layout that can be manufactured, and it includes tasks like clock tree synthesis, place and route, timing closure, and physical verification. 

Due to the distinct characteristics of the front-end and back-end stages, the approaches employed for HT prevention vary between the two stages. Front-end engineers may deploy different obfuscation techniques, such as logic locking \cite{35,33}, to protect the Intelectual Property (IP) of the design. In these methods, the design is secured with additional keys, initiating normal operation only when the correct sequence of keys is applied. While these techniques were not developed for HT prevention, it is understood that obfuscating the design against IP theft perhaps makes the design less evident for an adversary attempting to insert an HT.

Conversely, prevention methods applied in the back-end stage focus on protecting the chip layout from potential attackers within the foundry. One approach is layout filling which is aimed at restricting available resources, such as gaps and free routing tracks, to prevent adversaries from inserting malicious logic \cite{bisa, 40,papa}. Another strategy, known as split manufacturing, involves fabricating one part of the chip in one foundry and the remainder in a second foundry \cite{tigo}. Despite its promise, practical challenges arise, including finding two companies with compatible manufacturing technology and managing complex handshakes between the foundries. 

In the domain of DfHT, another research track focuses on developing methods to enhance HT detection \cite{54,57}. For this purpose, different types of checkers can be integrated into the design to sense irregularities and raise awareness in case of incorrect behavior. These checkers contribute to side-channel analysis by serving as thermal sensors to magnify thermal activity, or act as security checkers by introducing redundancy to the design. This approach aligns with strategies proposed for enhancing reliability against faults. In some works, such as \cite{trojanLessons}, alternative methods like functional testing and side-channel analysis are also considered for facilitating HT detection. However, in our classification, we view side-channel analysis and functional testing as primary methods for HT detection rather than simply facilitating detection. 

\begin{figure*}[!]
\begin{equation} \label{eq_nodes}
    \forall ~n \in \left \{ nodes_{(Des)} \right \}, ~n= \begin{cases}
C & \text{ if } functional~path ~exists ~between ~(n)~and ~(assertion) \\ 
V & \text{ if } functional~path ~does ~not~exist ~between ~(n)~and ~(assertion) 
\end{cases}
\end{equation}
\end{figure*}

\subsection{Threat Model}

The chip production process involves several stages, as shown in Fig. \ref{fig:threat_model}. Front-end engineers transform the high-level design description into a gate-level netlist through logic synthesis. This netlist is then handed to the back-end team, where engineers adjust it based on specific constraints such as area, power, and timing. Defensive techniques can be incorporated into the design by either the front-end or back-end team to protect against potential threats. The resulting layout is sent to a foundry for fabrication. After the chip is received, specific tests are performed on it to verify its functionality. 

In our threat model, we assume the foundry to be an untrusted facility, where a potential adversary (e.g., a rogue engineer) may be present. This encompasses fabrication-time attacks, similar to those explored in prior research, where malicious modifications to the design are introduced during the IC manufacturing \cite{icas}. Conceptually, a malicious foundry can incorporate three types of HTs into the layout of an IC: additive, substitution, and subtractive HTs. Additive HTs involve introducing additional circuit components and/or wiring into the existing design. Substitution HTs require removing logic that is replaced by HT circuit components and/or wiring. Subtractive HTs involve removing circuit components and/or wiring to alter the behavior of the existing circuit design. In this work, we \emph{only} focus on additive HT attacks due to their significant impact on system behavior, their detectability through changes in various design characteristics, and the extensive body of research on this HT type.
Design and test stages, including the design house and test house, are assumed to be trusted. Front-end and back-end teams implement defensive techniques in the design before sending it for fabrication, aiming to counter fabrication-time attacks.

Furthermore, we assume the attacker within the foundry has the capability to insert sophisticated, small, and rarely activated HTs that can evade side-channel analysis, logic testing, and simple forms of chip inspection. The attacker, due to his/her location, has access to the target technology's Process Design Kit (PDK) and advanced commercial CAD tools. Specifically, we focus on functional HTs that alter the chip's functionality, allowing their effects to be observed by comparing internal signals with the expected ones.

\subsection{Related Works}

\subsubsection{Logic Locking}
As mentioned before, logic locking is a key-based technique used to protect the intellectual property of ICs by obfuscating their functionality. However, this technique has limitations \cite{ll_survey} and has been subject to numerous attacks that can discover the key value. Attacks such as boolean Satisfiability (SAT) attacks, removal attacks, and bypass attacks, have demonstrated that logic locking can be vulnerable, allowing attackers to uncover the key and compromise the security of the design. A logic locking technique that relies on a secret key to control circuit functionality is presented in \cite{35}. The secret key becomes crucial for proper circuit operation, serving as an authentication mechanism. Another methodology is presented in \cite{33} to protect gate-level IPs, providing obfuscation and authentication. This approach modifies the state-transition function and internal-circuit structure, allowing normal operation with a predefined enabling sequence or key. However, it may have limitations against structural analysis-based attacks as it does not explicitly modify the state space of the existing Finite State Machine (FSM) \cite{33}.

\subsubsection{Layout Filling}

In \cite{bisa}, the authors propose a method of populating unused spaces with functional cells, creating an independent combinational circuit for post-fabrication testing against cell modifications. Nevertheless, a challenge lies in achieving a high occupation ratio while keeping the design routable. To overcome this limitation, \cite{papa} gives priority to filling gaps that could potentially be exploited for HT insertion. However, this strategy may lead to alterations in the initial routing, posing a risk of violating critical paths due to rerouting.

Addressing the limitation of user control over placement, \cite{ASSURER} proposes a placing refinement technique in which large unused layout spaces are segmented. Despite this refinement, there remains a vulnerability where attackers can reverse the refinement, creating optimal zones for their malicious logic. In a different approach presented in \cite{JohannICCAD}, a selective placement method is employed, where sensitive logic is positioned in denser layout areas, leading to places gaps around less sensitive regions.

Another attempt to mitigate layout vulnerabilities is discussed in \cite{gdsguard}, where the authors aim to reduce large gaps by shifting cells and utilizing the ECO features of commercial CAD tools. However, this technique often results in worsened timing, leading to negative timing slack in most cases.

\subsubsection{Online Monitors}
These techniques have been used widely for reliability and dependability, focusing on Concurrent Error Detection (CED) techniques, which introduce redundancy through parity codes or hardware duplication, complemented by a dedicated checker \cite{76}. 
The approach in \cite{57} focuses on combining special CED techniques and selective programmability to protect digital systems against HT attacks. However, this method imposes considerable area and power overheads on the design.



\section{Reusing Verification Assertions as HT detector} \label{sec:reuse} 
The foundation of this work originates from the utilization of assertions as HT detectors, as discussed in \cite{assertion}. A hardware assertion is a statement or condition commonly specified in HDLs that defines a certain expected behavior or property of a digital circuit. This concept gains significance, particularly when extensive time and effort are invested in the verification of designs to ensure their integrity and the absence of bugs. The generated verification assets (i.e., assertions), which are often no longer utilized after the verification process, can be repurposed for security applications, specifically for the detection of HTs. The main challenge lies in the absence of a technique to synthesize the assertions as checkers, as well as a metric that identifies suitable assertions for use as HT detectors. To address this gap, we introduced a novel metric termed Security Coverage (SC) that evaluates the security effectiveness of verification assertions \cite{assertion}.

\subsection{Security Coverage}
As the name indicates, SC focuses on assessing the security qualifications of existing assertions. It is defined as the ratio of nodes covered by the assertion to the total nodes within the design. Each node \((n)\) in the design is defined as either covered \((C)\) or vulnerable \((V)\), as indicated in Eq. \ref{eq_nodes}. 

Therefore, the SC of a design is calculated as follows: 

\begin{equation} \label{eq_sc}
    SC_{(Des)}=\frac{|\bigcup_{i=1}^{k}C_{i}|}{|\bigcup_{i=1}^{k}C_{i}|+|\bigcup_{i=1}^{k}V_{i}|}
\end{equation}

Where \textit{k} denotes the total number of the assertions, and \(C_{i}\) and \(V_{i}\) are the covered and vulnerable nodes, respectively. 


An illustrative example representing a design with two integrated assertions is presented in Fig.~\ref{fig:assr_example}. To calculate the SC for the entire design, Eq. \ref{eq_sc} is applied. Furthermore, this equation can be employed to determine the SC for each individual assertion, enabling a comparative analysis of their security properties. Considering the two assertions in this scenario, the SC for each assertion is computed by finding the number of covered nodes \((C)\). Examining Fig.~\ref{fig:assr_example}\subref{assr_1}, nodes 1, 2, 3, 6, 7, 9, 12, 13, and 16 (highlighted in green) have at least one functional path\footnote{A functional path refers to a path within a circuit that can be traversed using a combination of valid inputs. This contrasts with non-functional paths that are unreachable with a given set of inputs.} to Assr\_1. Similarly, the covered elements for Assr\_2 (highlighted in blue) encompass nodes 2, 3, 4, 5, 7, 8, 10, and 17, as depicted in Fig.~\ref{fig:assr_example}\subref{assr_2}. It is worth noting that certain nodes may appear in multiple subsets of covered nodes for different assertions, exemplified by nodes 2, 3, and 7, while some other nodes may not be covered by any assertions (nodes 11, 14, 15, and 18).

\begin{figure*}[ht]
    \centering
    \subfloat[]{\includegraphics[height=.35\columnwidth]{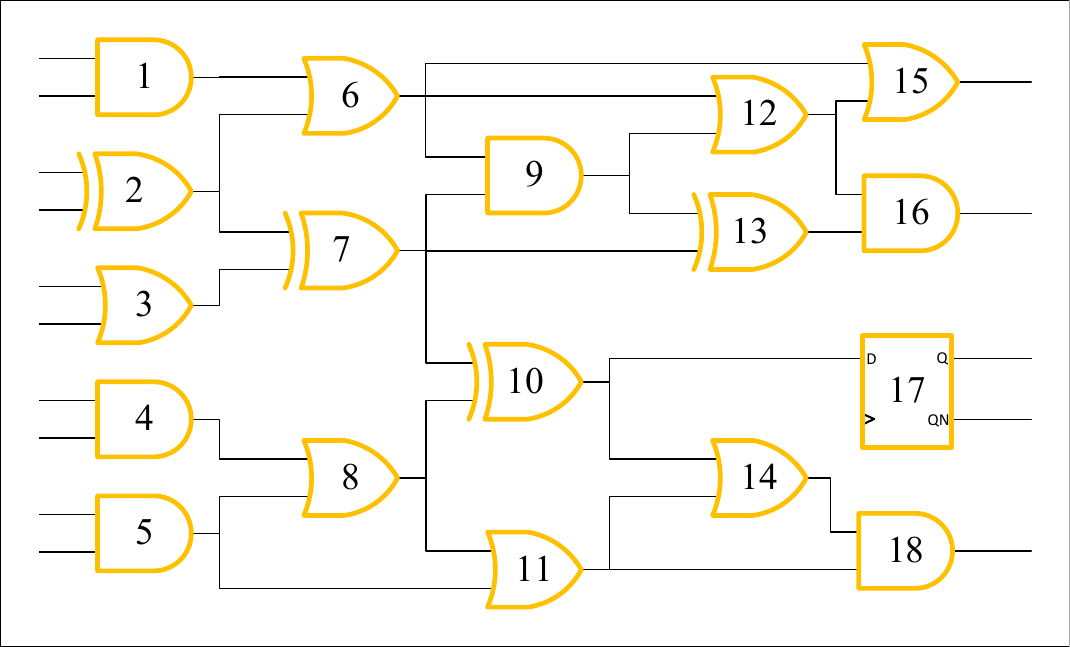} \label{assr_full}} \hspace{1mm}
    \subfloat[]{\includegraphics[height=.35\columnwidth]{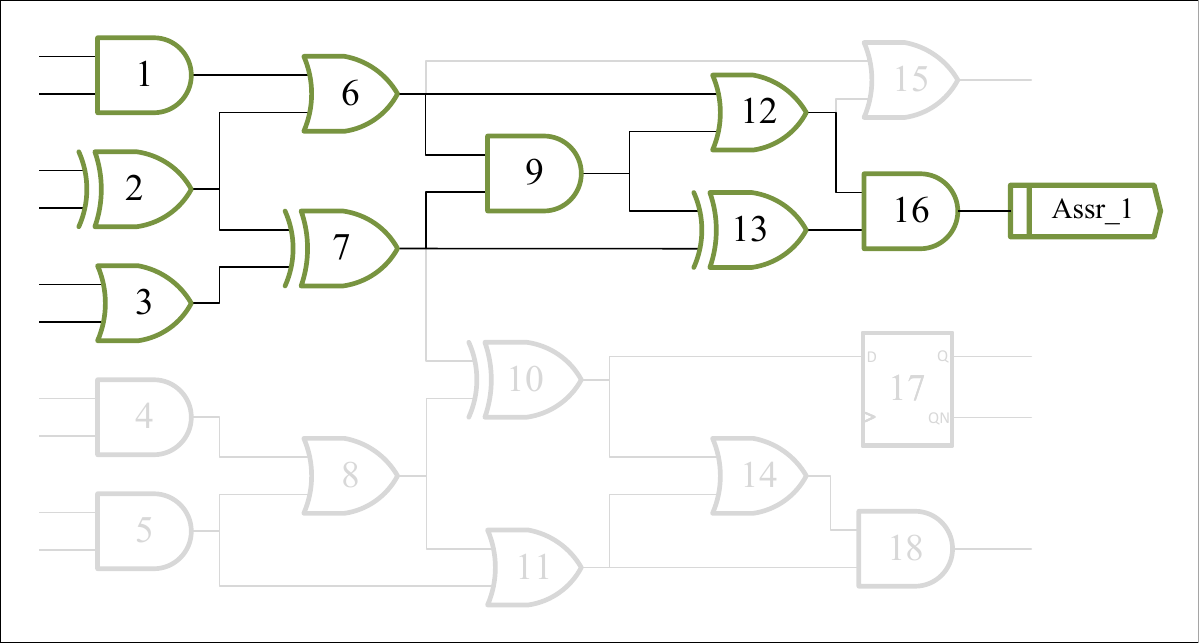} \label{assr_1}} \hspace{1mm}
    \subfloat[]{\includegraphics[height=.35\columnwidth]{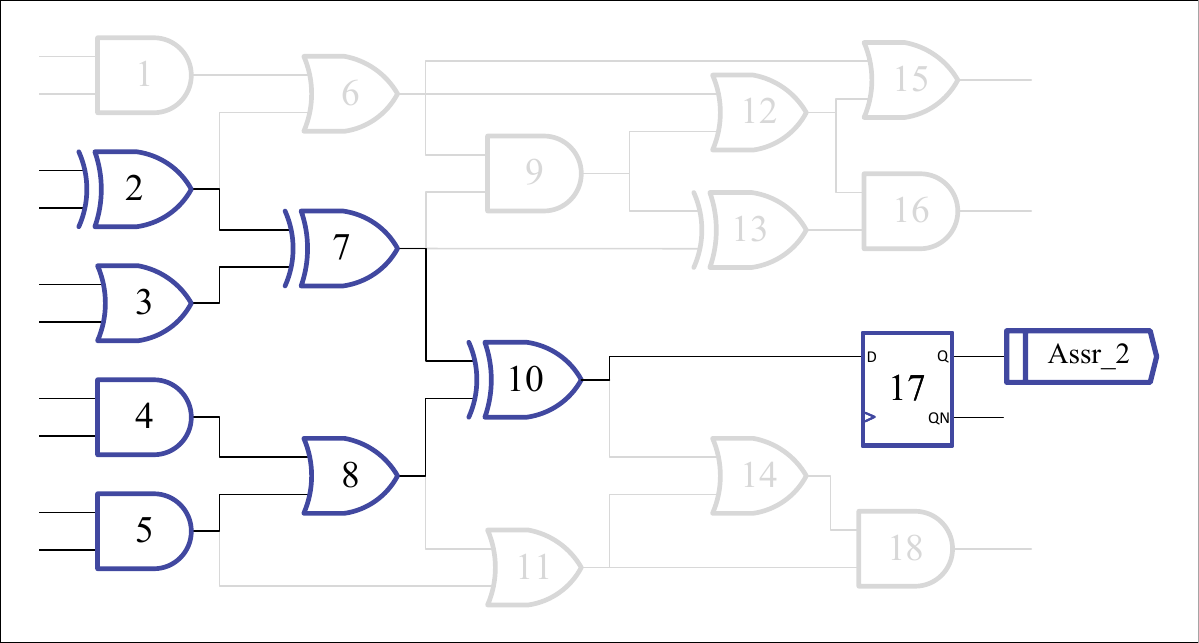} \label{assr_2}}
    \caption{An example of a) original design, b) nodes covered by bound assertion Assr\_1, and c) nodes covered by bound assertion Assr\_2}
    \label{fig:assr_example}
\end{figure*}

Nevertheless, several challenges restrict the calculation and use of SC for assertions in a design:

\begin{enumerate}
    \item \textbf{Synthesis Limitations:} The majority of verification assertions are written to identify irregularities during simulation, and they cannot be synthesized. Consequently, they cannot be bound to the design.
    \item \textbf{Functional Path Requirement:} To consider a node as covered, there must be a functional path between the node and the output of the assertion. Hence, the existence of a connection between the node and the output of the assertion is not sufficient, and traditional methods, such as extracting the input cone(s), cannot be used. 
    \item \textbf{Overheads:} It is crucial to acknowledge that individual assertions introduce specific overheads to the design. It is also the case that combined assertions introduce overheads that are not the sum of their individual overheads. In evaluating the effectiveness of each assertion, it is necessary to consider these factors to allow for a comprehensive evaluation of the trade-off space between security and the associated design overheads.
\end{enumerate}

The synthesis limitations problem is the main challenge since it affects other problems. The most widely used languages for describing assertions are PSL and SystemVerilog, but the use of standardized languages does not imply that they are synthesizable -- some assertions are clearly only meant for simulation purposes and have to be filtered out. To address this, we utilize the MBAC tool~\cite{mbac} to convert PSL and/or SystemVerilog assertions into a synthesizable Verilog format.

Once the assertions are transformed into synthesizable code, they are integrated with the main circuit for an assessment of their effectiveness and the overhead they introduce. To address the second challenge and obtain the SC for assertions, we use the Cadence JasperGold Security Path Verification (SPV) tool \cite{jaspergold}. This tool allows us to perform a proof analysis for verifying the existence of functional paths between desired nodes in the design, also known as taint analysis. It should be noted that JasperGold could have been replaced by any other tool capable of doing taint analysis, academic or commercial. In our scenario, the considered origin nodes are all nodes in the design; the destination node is the assertion output. The existence of a functional path between any node in the design (origin) and the assertion output (destination) means that the node is covered by the assertion (see Fig. \ref{fig:assr_example}). 

In other words, if a covered node's value is maliciously manipulated (e.g., due to an inserted HT activating its payload), the assertion can detect it, functioning as an online monitor. This concept shares similarities with verification schemes, where assertions aim to identify irregularities in the design. However, the key difference is that verification assertions are only necessary until the logical synthesis step. Verification engineers mainly seek to correct the potential mistakes made by the design team, and they can achieve their objectives through simulation, ensuring the functionality of the design. Once they confirm that the design is bug-free, assertions are no longer needed since the design's functionality and specification remain constant in subsequent chip design steps. In contrast, security engineers must consider potential threats during fabrication, and assertions that have transformed into online monitors must remain with the design until the chip is fabricated.

The third challenge involves quantifying the overheads introduced by each assertion on the design. While simulation offers insights into the circuit's incorrect behavior and internal values, it falls short in determining crucial performance characteristics like power, timing, and area. To address this limitation, the design undergoes multiple synthesis runs, and precise performance reports are generated. Initially, the original circuit is synthesized without the assertions, providing maximum clock frequency, power, and area reports. Subsequently, the circuit, now integrated with the bound assertions functioning as embedded security checkers, undergoes synthesis again. The evaluation is based on a comparative analysis of results from the two synthesis processes, ensuring a comprehensive assessment of assertion performance in terms of design overhead.

Once the SC and the overheads of each assertion are known, we can decide if the assertion is worth to be kept or not. However, this cannot be performed manually since investigating the trade-off between SC and overheads for each assertion is a time-consuming task. Hence, an automation flow is needed to select the assertions. The prerequisite of automation flow is defining a strategy to only pick efficient assertions in terms of security and overheads. This efficiency can be defined such that the assertion has more security properties, imposes less overheads on the design, or a balance between both.

Fig.~\ref{fig:flow} illustrates a proposed automated flow for the assertion selection process. The initial step involves selecting an assertion from the candidate list, which contains assertions with synthesizable potential, excluding those with a simulation-based nature that have already been filtered out. The chosen assertion is then converted into synthesizable logic and integrated into the design. This prepared design enables overhead evaluation, where synthesis is performed with the assertion to compare various metrics (e.g., power, performance, and area (PPA)) against the values obtained from the original design.


\tikzstyle{shape0} = [rectangle, rounded corners, minimum width=3cm, minimum height=1cm, text centered, font=\large, color=color19, draw=color18, line width=1, fill=color4]
\tikzstyle{shape1} = [trapezium, trapezium left angle=70, trapezium right angle=110, minimum width=6cm, minimum height=1cm, text centered, font=\large, color=color3, draw=color2, line width=1, fill=color1]
\tikzstyle{shape2} = [rectangle, minimum width=5cm, minimum height=1.5cm, text centered, font=\large, color=color3, draw=color5, line width=1, fill=color4]
\tikzstyle{shape3} = [rectangle, minimum width=5cm, minimum height=1cm, text centered, font=\large, color=color3, draw=color5, line width=1, fill=color4]
\tikzstyle{shape4} = [rectangle, minimum width=5cm, minimum height=1cm, text centered, font=\large, color=color3, draw=color9, line width=1, fill=color8]
\tikzstyle{shape5} = [diamond, minimum width=1cm, minimum height=1cm, font=\normalsize, color=color3, draw=color7, line width=1, fill=color10]
\tikzstyle{shape6} = [diamond, minimum width=1cm, minimum height=1cm, font=\normalsize, color=color3, draw=color12, line width=1, fill=color4]
\tikzstyle{shape7} = [rectangle, minimum width=1cm, minimum height=1cm, text centered, font=\large, color=color3, draw=color7, line width=1, fill=color13]
\tikzstyle{shape8} = = [thick, draw=color16, line width=2, >->, >=stealth]
\tikzstyle{shape9} = [thick, draw=color16, line width=2, ->, >=stealth]

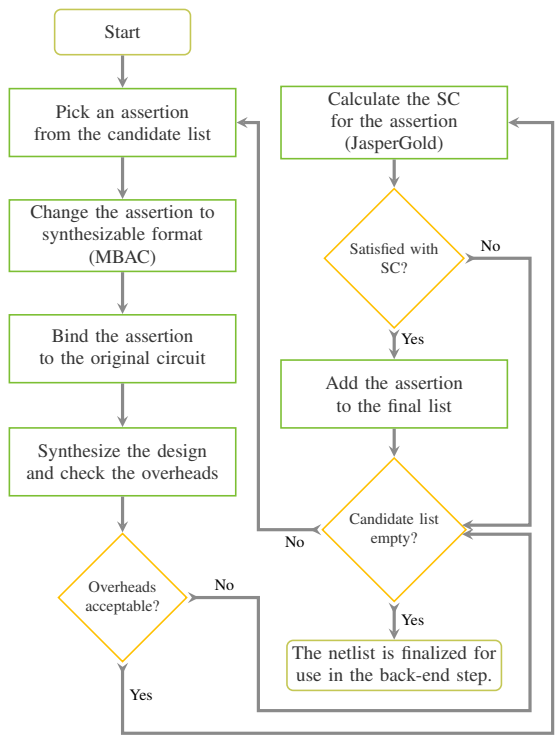
\begin{figure}[!]
\centering
\begin{adjustbox}{width=0.4\textwidth}
\begin{tikzpicture}[node distance=2cm, on grid, auto]
\node (obj1) [shape0]                                                             {Start};
\node (obj2) [shape2, below of=obj1, align=center]                                {Pick an assertion \\ from the candidate list};
\node (obj3) [shape2, right of=obj2, xshift=4cm, align=center]                    {Calculate the SC \\ for the assertion \\ (JasperGold)};
\node (obj4) [shape2, below of=obj2, yshift=-.5cm, align=center]                  {Change the assertion to \\ synthesizable format \\ (MBAC)};
\node (obj5) [shape6, below of=obj3,  yshift=-1cm, align=center]                  {Satisfied with \\ SC?};
\node (obj6) [shape2, below of=obj4, yshift=-.5cm, align=center]                  {Bind the assertion \\ to the original circuit };
\node (obj7) [shape2, below of=obj5, yshift=-1cm, align=center]                   {Add the assertion \\ to the final list};
\node (obj8) [shape2, below of=obj6, yshift=-.5cm, align=center]                  {Synthesize the design \\ and check the overheads };
\node (obj10) [shape6, below of=obj7, yshift=-1cm, align=center]                  {Candidate list \\ empty?};
\node (obj11) [shape6, below of=obj8, yshift=-1cm, align=center]                  {Overheads \\ acceptable?};
\node (obj16) [shape0, below of=obj10, yshift=-1cm, align=center]                               {The netlist is finalized for \\ use in the back-end step.};
\draw [shape8] (obj11) -- node[pos=.45]{Yes} ++(0,-3) -| ++(9.5,13.5)  -- (obj3)  ;
\draw [shape8] (obj11) -- node[pos=1.45]{No} ++(2,0) -| ++(1,-2.5)  -| ++(6,0)  -| ++(0,3.9) -- ++(-1.5,0)  ;
\draw [shape9] (obj1) --  (obj2);
\draw [shape9] (obj2) --  (obj4);
\draw [shape9] (obj4) --  (obj6);
\draw [shape9] (obj6) --  (obj8);
\draw [shape9] (obj8) --  (obj11);
\draw [shape9] (obj3) --  (obj5);
\draw [shape8] (obj5) -- node[pos=.5]{Yes} ++(0,-2) -- (obj7);
\draw [shape8] (obj5) -- node[pos=1.3]{No} ++(2,0) -| ++(1,-5.9) -- ++(-1.5,0);
\draw [shape9] (obj7) --  (obj10);
\draw [shape8] (obj10) -- node[pos=1.5]{No} ++(-2,0) -| ++(-1,9)  -- (obj2)  ;
\draw [shape8] (obj10) -- node[pos=1]{Yes} ++(0,-2) -- (obj16);
\end{tikzpicture}
\end{adjustbox}
 \vspace{-2mm}
\caption{Optimization flow for selecting the assertions to be used as security checkers}
\label{fig:flow}
\end{figure}

After assessing the overheads and confirming their compatibility with user-defined requirements, the subsequent step involves calculating the SC. This is achieved by employing Eq. \ref{eq_sc} and utilizing the JasperGold SPV tool. If the assertion passes both the overhead evaluation and achieves the desired SC, it is included in the final list of assertions. Otherwise (e.g., if the overheads are regarded as unacceptable or the SC falls short of the user's expectations), an alternative assertion from the candidate list is considered (if any). It should be noted that the decision in this step can be based on either the SC of the assertion individually or the overall SC threshold, which is specified by the user for the whole design. Once the flow is completed, the generated netlist with the embedded assertion(s) is considered finalized. This finalized netlist will be used in the back-end stage for further security improvements through the insertion of online monitors.

\subsection{Evaluation of the Security Properties of Verification Assertions}
In a practical case study assessing the security properties of verification assertions, we conducted an SC calculation for predefined assertions in various IPs of the OpenTitan System on Chip (SoC). 
Given that OpenTitan, an open-source silicon root of trust project, is developed and maintained by a community of experienced engineers, it is recognized as a highly reliable and suitable case study for evaluating the effectiveness of our proposed framework. The initial experiment focused on selecting four distinct assertions present in 35 different Register Top modules of OpenTitan, resulting in a set of 140 assertions. This initial analysis indicates a selection based on commonality across different IPs, disregarding specific characteristics or overheads. 

The SC of each assertion in different IPs are depicted in Fig~\ref{fig:opentitan}. As shown in this figure, the calculated SC percentages for each assertion are less than 5\%, which may raise concerns, even when overlooking the associated overheads. However, it is crucial to recognize that a lower SC does not necessarily imply inadequate security properties, as some applications prioritize safeguarding specific security-sensitive components rather than the entire design.

\begin{figure*}[ht]
    \centering
    \includegraphics[width=.95\textwidth, trim=.1cm 8.5cm .1cm 7.5cm, clip]{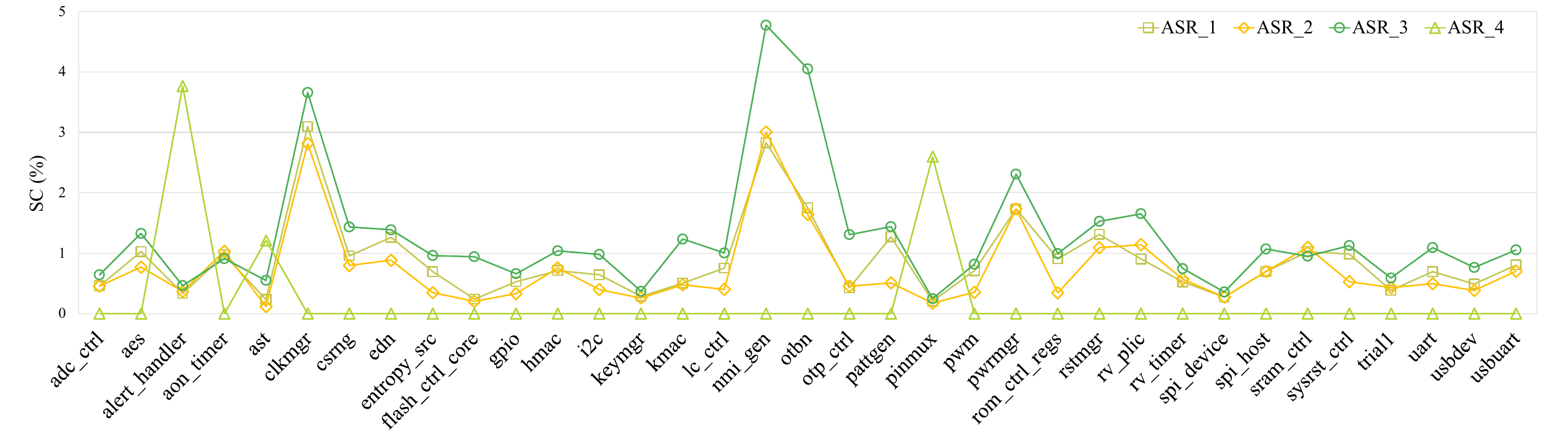}
    \vspace{-2mm}
    \caption{SC percentage for different IPs of OpenTitan}
    \label{fig:opentitan}
\end{figure*}


The SC of an assertion is influenced by various factors, with a crucial one being the nature of how the assertion is formulated for verification. Specifically, if an assertion is defined to concentrate only on local signals, such as checking whether specific bits of a register are 0 or 1, it is generally considered less impactful. On the contrary, assertions that are not restricted to narrow scopes and instead describe high-level behaviors are preferable. Consequently, for users prioritizing a higher SC for the entire design over assertions safeguarding specific design regions, the emphasis should be on selecting assertions covering larger design parts. The initial experiment did not prioritize assertions based on high-level descriptions but focused on finding synthesizable assertions present in multiple IPs. However, as depicted in Fig.~\ref{fig:opentitan}, the SC results are generally unremarkable. This underscores the need for further analysis when choosing assertions, rather than blind selection.

In the subsequent experiment, we investigate the SC of assertions uniquely generated for deployment in specific IPs, causing them to be unsuitable for use in other IPs. This experiment demands more effort, as synthesizable assertions undergo a preliminary analysis before binding into the design. Consequently, not only unsynthesizable assertions are excluded from consideration, but also assertions focused on checking local signals are filtered out.

Figure~\ref{fig:sc_ips} illustrates the calculated SC percentages for individual assertions within three chosen IPs. As shown in this figure, the SC for each assertion is notably higher compared to the figures obtained from the initial experiment (Fig.~\ref{fig:opentitan}). The average SC stands at 85.83\% for the selected assertions in the \textsc{alert\_handler}  (Fig.~\ref{fig:sc_ips}\subref{alert_sc}), 46.03\% for the selected assertions in the \textsc{alert\_handler\_esc}  (Figure~\ref{fig:sc_ips}\subref{alert_esc_sc}), and 38.35\% for the selected assertions in the \textsc{flash\_phy\_rd} module (Figure~\ref{fig:sc_ips}\subref{flash_sc}).

\begin{figure*}[ht]
    \centering
    \subfloat[]{\includegraphics[width=.67\columnwidth, trim=3.5cm 4cm 1.9cm 3.9cm, clip]{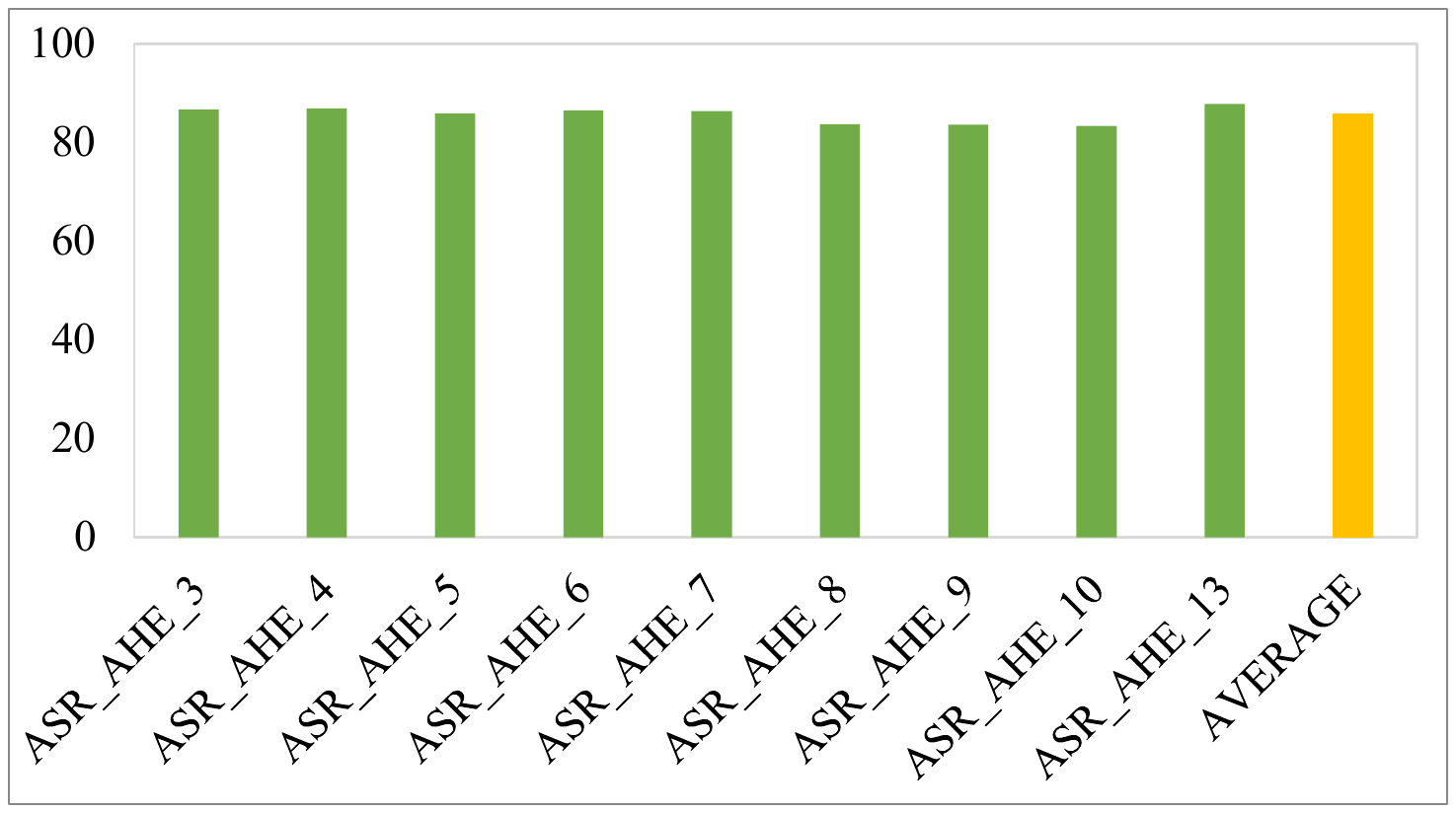}\label{alert_sc}} 
    \subfloat[]{\includegraphics[width=.67\columnwidth, trim=3.3cm 4cm 2.2cm 3.9cm, clip]{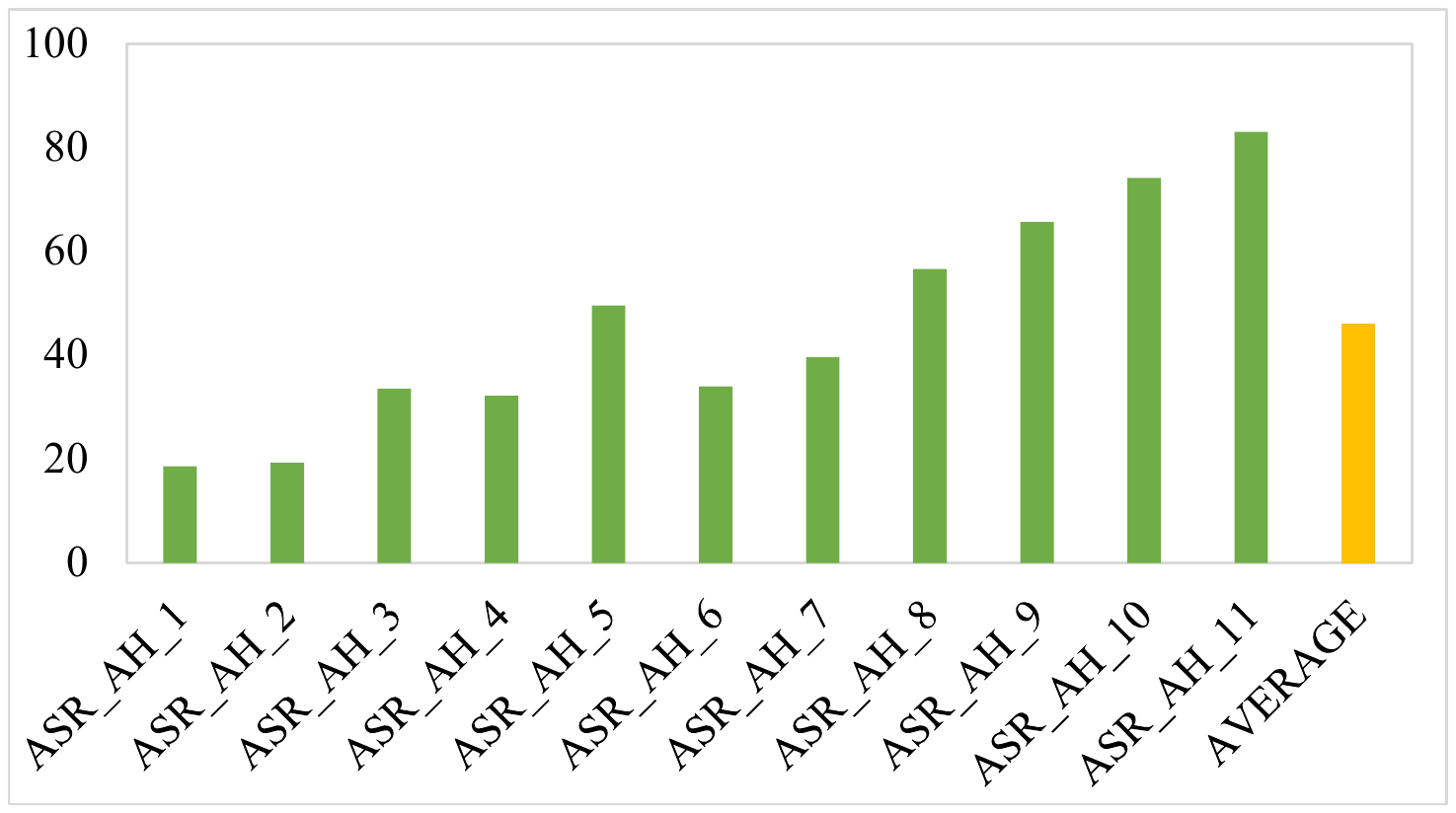} \label{alert_esc_sc}}
    \subfloat[]{\includegraphics[width=.67\columnwidth, trim=3.5cm 4cm 2cm 3.9cm, clip]{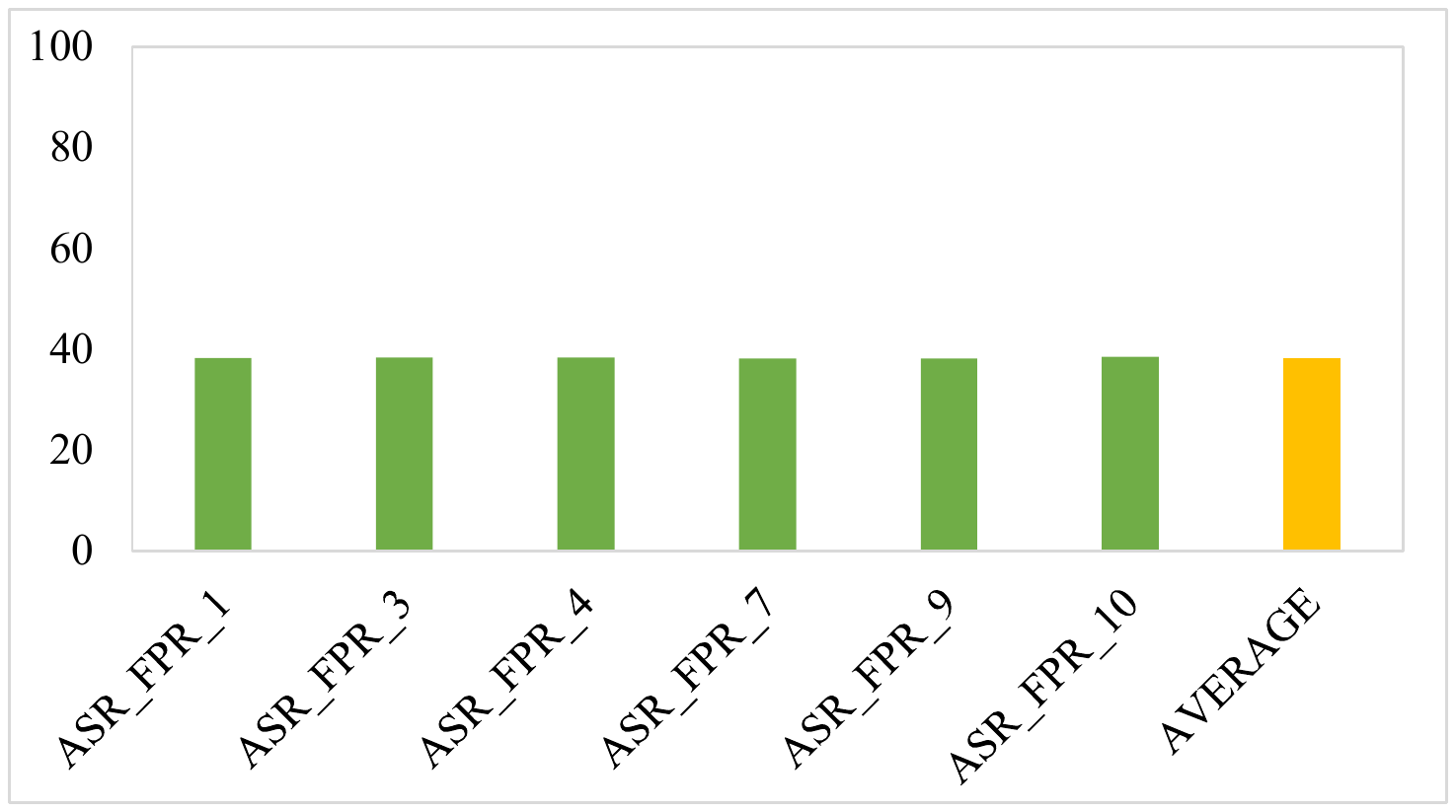} \label{flash_sc}} 
    \vspace{-1mm}
    \caption{Calculated SC percentage for the different assertions in selected IPs of OpenTitan: a) alert\_handler\_esc, b) alert\_handler, and c) flash\_phy\_rd}
    \label{fig:sc_ips}
\end{figure*}

\subsection{Challenges}
While achieving higher SC numbers for different assertions (e.g., in the \textsc{alert\_handler\_esc} IP as shown in Fig.~\ref{fig:sc_ips}\subref{alert_sc}) may suggest improved design security, it is crucial to acknowledge that ensuring design security goes beyond relying solely on SC metrics, even with high percentages (e.g., exceeding 80\%). There are notable challenges associated with SC, and some key ones include:

\begin{enumerate}
    \item Ineffectiveness of verification assertions at runtime: While functional assertions can cover various security properties by ensuring the expected behavior of the circuit, they may lack the precision to thoroughly cover the negative or unexpected behavior of a circuit under attack. Although the SPV tool is employed to calculate taint propagation coverage, the effectiveness of detecting HT behavior by the synthesized assertions at runtime is not guaranteed.
    \item Scalability concerns: The necessity to ``bind the assertion" and synthesize the entire design for characterizing overheads may pose scalability challenges for larger designs. The resource-intensive nature of this process could limit its practicality for more extensive and complex circuits.
    \item Assertion availability: The methodology aims to reuse existing assertions rather than generating new ones. However, a potential challenge arises if no suitable assertion with acceptable SC is found for a given design. This situation is exemplified by the SC numbers obtained for different assertions, as depicted in Fig.~\ref{fig:opentitan}.
\end{enumerate}

To overcome these challenges, an additional layer of security must be incorporated into the design to address the limitations associated with reusing the verification assertions as online monitors. 

\section{Enhancing the Security by Adding Online Monitors during the Physical Synthesis}\label{sec:online_monitors}
In this section, we introduce a novel methodology to incorporate online monitors into the layout during the physical synthesis flow. While prior works have explored the idea of integrating online monitors, most efforts have concentrated on introducing these checkers during the front-end stage of design. In our work, we take a different approach by directly incorporating the checkers into the layout while considering front-end inserted assertions.

Figure~\ref{fig:layout} provides a simplified view of the layout of a block of an IC. The green polygons represent standard cells that are later interconnected through various metal layers, establishing the logical function of the design. Due to fabrication complexities, particularly in modern process nodes, achieving 100\% density where the layout is entirely filled with standard cells, is impractical. Thus, gaps are present in the layout, highlighted in red in Fig.~\ref{fig:layout}. These gaps can be later used by an adversary for inserting his/her malicious logic (i.e., HTs) \cite{bisa}.

Despite these gaps being filled with filler cells before being sent to fabrication, since these filler cells lack functionality and are not connected to the original design, they can be easily removed by the attacker inside the foundry. Our methodology leverages these gaps and available resources to insert online checkers into the design. By doing this, not only do we add an extra security layer to the design, but also we limit the adversary to put malicious logic by increasing the density and congestion in the layout.

\begin{figure}[htbp]
    \centering
    \includegraphics[width=.37\textwidth, trim=.1cm .5cm .1cm .5cm, clip]{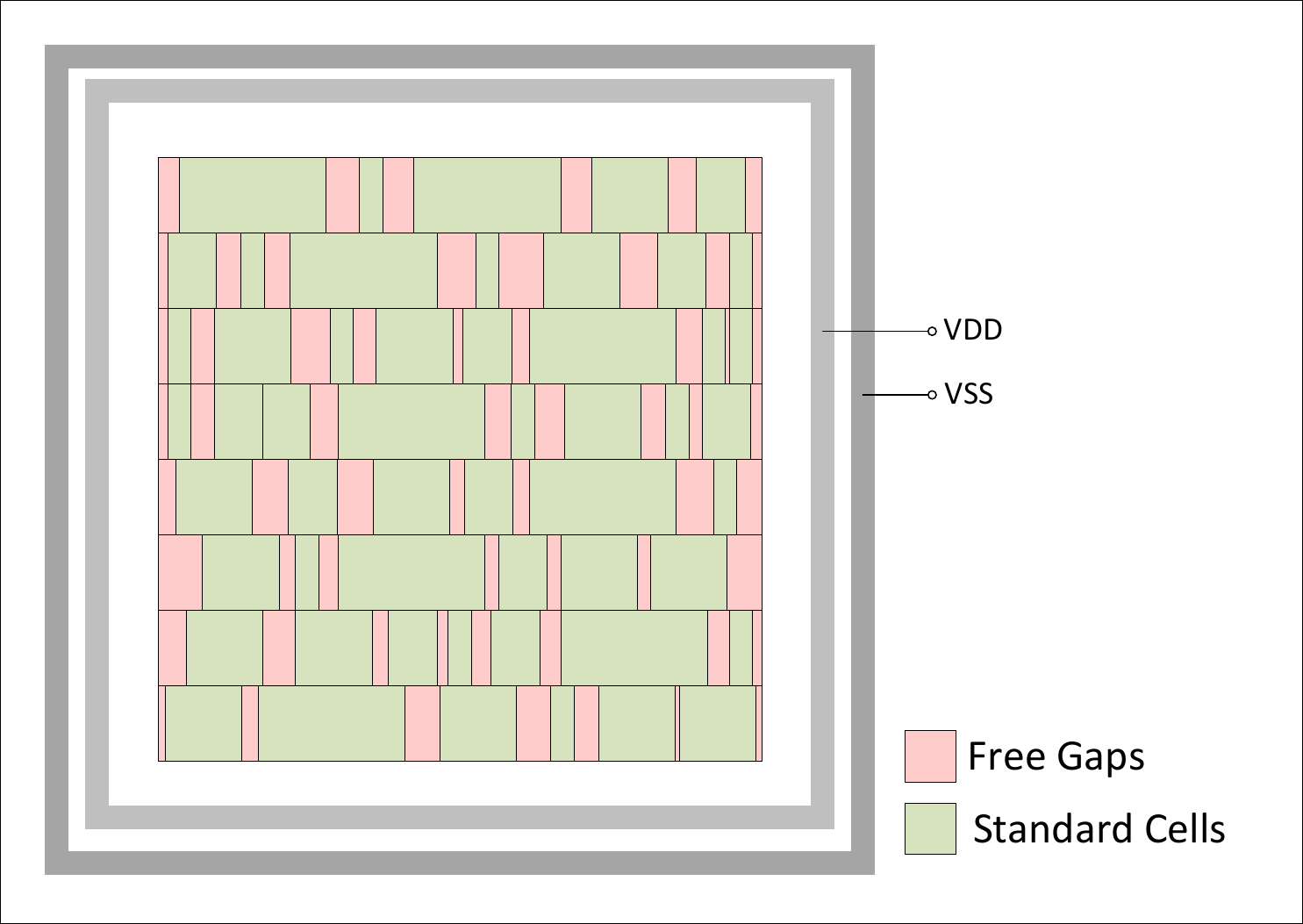}
    \vspace{-2mm}
    \caption{Illustrative example of a block layout within a chip}
    \label{fig:layout}
\end{figure}

Although this methodology can also be used independently with minor modifications in the flow, we use it as a complementary methodology along with reusing the assertion for detecting HTs to cover its shortcomings. It should be noted that we leverage the ECO features of the CAD tools for inserting the online monitors into the layout\footnote{ECO features in CAD tools allow engineers to make last-minute modifications to the existing layout, such as adding or removing components, changing connections, etc. These changes may be necessary due to factors such as updated specifications, errors discovered in the initial design, or other requirements that emerge during the design phase.}. This use of ECO minimizes alterations to the original layout with each added online monitor, ensuring optimal overheads in comparison to front-end approaches.

\subsection{Online Monitors}
As mentioned earlier, the online monitors act as an extra protection layer for the nodes not covered by assertions. To create this protection strategy, we use a Dual Modular Redundancy (DMR) scheme, as shown in Fig.~\ref{fig:onlineChecker}. The left image (Fig.~\ref{fig:onlineChecker}\subref{original}) illustrates a subpart of a design represented in Fig.~\ref{fig:assr_example}, in which the covered nodes (10 and 17) are highlighted in green, whereas the vulnerable/uncovered nodes (11, 14, and 18) are highlighted in red. To construct an online monitor for this subset of the design, first, we duplicate the uncovered gates with exactly the same equivalent gates from the library (i.e., the same gate type, and same drive strength). Then, we compare the output of the duplicated part with the output of the original part by XORing these two signals, as depicted in Fig.~\ref{fig:onlineChecker}\subref{redundant}. Therefore, adding an online monitor for this part adds seven new covered nodes (11, D11, 14, D14, 18, D18, and V18) to the previously covered nodes (10 and 17). 

It is important to highlight that an adversary may attempt to substitute the online monitors by a HT. However, the online monitors are always on and therefore contribute to the power profile of the circuit. An HT, on the other hand, should have a stealthy trigger, which is not compatible with an always-on functionality. An HT that promotes a noticeable change in the chip's power profile is likely detectable.

\begin{figure}[htbp]
    \centering
    \subfloat[]{\includegraphics[width=.42\columnwidth, trim=.2cm .4cm .2cm .2cm, clip]{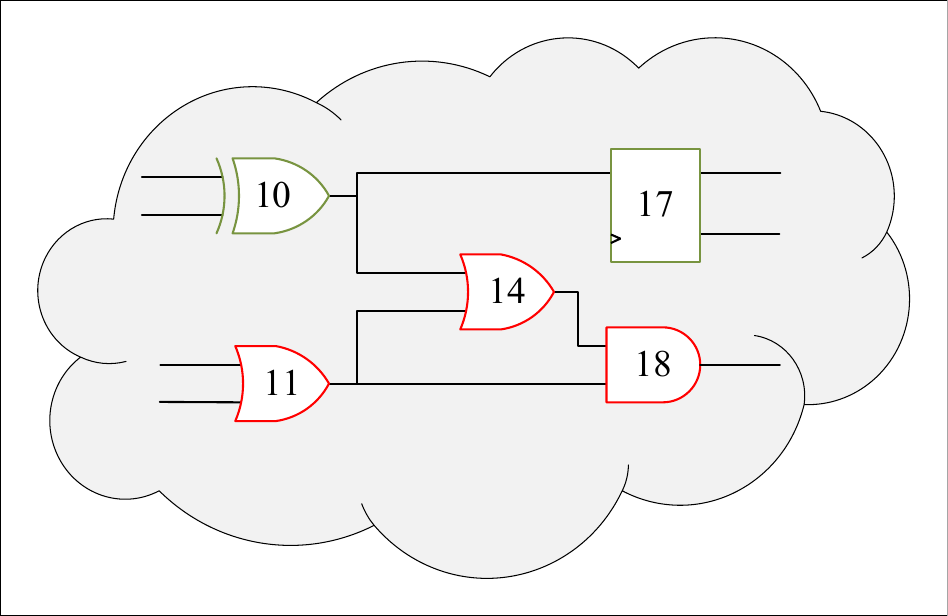}\label{original}} 
    \subfloat[]{\includegraphics[width=.55\columnwidth, trim=1cm .4cm .2cm .6cm, clip]{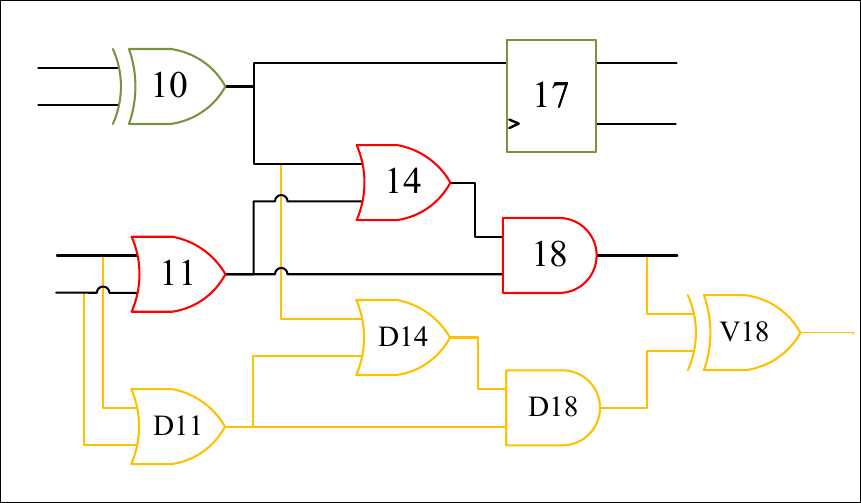} \label{redundant}}
    \caption{An example of a) design before adding online monitors, and b) design with the protection logic (D11, D14, and D18 as the duplicates and V18 as the voter) to protect the uncovered gates (11, 14, and 18)}
    \label{fig:onlineChecker}
\end{figure}

A greater number of connected uncovered gates is desirable, as all of these connected gates require only one XOR gate as a voter. 
However, limitations inherent in the physical synthesis flow may pose challenges to implementing DMR for all uncovered gates, even if they form a cone, as illustrated in Fig.~\ref{fig:onlineChecker}. Further details on this aspect are elaborated in the subsequent section.  

\subsection{Embedding Online Monitors into the Layout}
The comprehensive flow for integrating online monitors during the physical synthesis flow is outlined in Fig.~\ref{fig:backend_flow}. It involves four major steps that are detailed as follows:

\tikzstyle{shape10} = [tape, minimum width=2.2cm,  tape bend top=none, minimum height=1.5cm, text centered, font=\large, color=color3, draw=color12, line width=1, fill=color4]
\tikzstyle{shape11} = [rectangle, minimum width=4cm, minimum height=1.5cm, text centered, font=\large, color=color3, draw=color5, line width=1, fill=color4]
\tikzstyle{shape12} = [rectangle, minimum width=6cm, minimum height=1.5cm, text centered, font=\large, color=color3, draw=color5, line width=1, fill=color4]

\begin{figure}[htbp]
  \centering
  \begin{adjustbox}{height=0.4\textheight}
    \begin{tikzpicture}[node distance=2cm, on grid, auto]

      \node (netlist) [shape10, align=center] {Netlist};
      \node (spv_reports) [shape10, right=3.2cm of netlist, align=center] {Uncovered \\ Nodes};
      \node (pnr) [shape11, below=of netlist, xshift=.8cm, yshift=-.3cm, align=center] {Physical Synthesis \tikz \node[draw, circle, inner sep=1pt , minimum size=1.5mm] {1};};
      \node (layout) [shape10, below=of pnr, yshift=-.3cm, align=center]  {Layout};
      \node (density) [shape12, below=of layout, yshift=-.3cm, xshift=1cm, align=center] { Density Analysis for the \\ Uncovered Nodes \tikz \node[draw, circle, inner sep=1pt , minimum size=1.5mm] {2};};
      \node (fanin) [shape12, below=of density, align=center] {Ranking the Candidates Based \\on the Fanin Cone Size \tikz \node[draw, circle, inner sep=1pt , minimum size=1.5mm] {3};};
      \node (eco) [shape12, below=of fanin, align=center] {Generating Layouts for \\ the ECO Round \tikz \node[draw, circle, inner sep=1pt , minimum size=1.5mm] {4};};
      \node (lays) [shape10, below=of eco, align=center] {};
      \node (lays1) [shape10, right=of lays, xshift=-1.8cm, yshift=-.1cm, align=center] {};
      \node (lays2) [shape10, right=of lays1, xshift=-1.8cm, yshift=-.1cm, align=center] {};
      \node (lays3) [shape10, right=of lays2, xshift=-1.8cm, yshift=-.1cm, align=center] {Protected \\ Layouts};

      \draw [shape9] (netlist) -- ++(0,-1.55);
      \draw [shape9] (spv_reports) -- ++(0,-6.15);
      \draw [shape9] (pnr) -- (layout);
      \draw [shape9] (layout) -- ++(0,-1.55);
      \draw [shape9] (density) -- (fanin);
      \draw [shape9] (fanin) -- (eco);
      \draw [shape9] (eco) -- (lays);

    \end{tikzpicture}
  \end{adjustbox}
  \vspace{-2mm}
  \caption{An overall flow of integrating online monitors during the physical synthesis}
  \label{fig:backend_flow}
\end{figure}
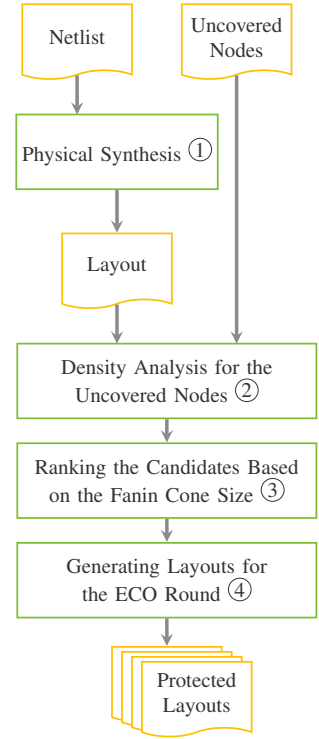

\subsubsection{Physical Synthesis} The first step starts with turning the netlist into a layout using a physical synthesis tool. Here, the netlist includes the main design and all selected assertions bound to it. This step involves several stages such as placement, clock tree synthesis, and routing. Through physical synthesis, vital information such as the precise placement of standard cells, the physical distribution of the clock, and the structure of interconnections in terms of wire length or the utilization of each available metal layer is obtained. Essentially, physical synthesis provides insight into the spatial configuration of the design. It should be noted that this time-consuming step is performed only once for each design.

\subsubsection{Density Analysis for the Uncovered Nodes} The resulting layout, along with the report of uncovered nodes from the SPV tool, becomes the input to our developed analytical tool. The objective here is to identify the uncovered nodes with available gaps around them, suitable for housing the online checker tasked with safeguarding the respective node. If there is no space available around the uncovered node, the online checker might be placed at a far distance from it, leading to higher resource utilization and degradation of the PPA parameters of the design. This is a crucial step in our flow to prevent this from happening in order to minimize layout modifications. In other words, the density analysis provides an adequate basis for embedding the online monitors for future ECO rounds. This makes the presented methodology different from front-end protection schemes where the actual overheads are often ignored. It should be noted that the scope of the searching area around each node is adjustable and can be changed according to the design size and density. A larger search area increases the processing time because more potential locations must be evaluated, and it can also lead to increased use of routing resources, which can negatively impact the overall performance of the design due to longer interconnect delays and higher power consumption. Therefore, the balance between the size of the search area and the associated overhead must be carefully managed to optimize both the detection capabilities and the design performance.

\subsubsection{Ranking the Candidates Based on the Fanin Cone Size} The generated candidates from the density analysis tool undergo a ranking process. As emphasized earlier, our preference is to place the online checker for a group of interconnected uncovered nodes rather than individual gates. This choice offers advantages in terms of area, power, and routing resources. Therefore, our ranking system assigns higher priority to subsets of candidate gates with larger input cones, optimizing the overall efficiency of the protection scheme. It should be noted that in our cone analysis, we only consider the candidates with a cone size of 2 or greater to improve the efficiency of our methodology. 

\subsubsection{Generating Layouts for the ECO Round} The final step involves the generation of protected layouts. To facilitate this, we developed a tool that takes the ranked list of candidates (generated in the previous step) and integrates the online checkers into the layout incrementally. More specifically, one online monitor is added to the design at a time, and a new layout containing the added monitor is generated. This process is repeated from the top of the ranked list to the end. Therefore, if there are \emph{n} candidates (the nodes suitable to be protected by online checkers) in the ranked list, we create \emph{n} different layout files in an iterative fashion. Each protected layout file, such as \emph{Layout1}, contains the online monitor from \emph{Candidate1}; \emph{Layout2} contains the online monitors for \emph{Candidate1} and \emph{Candidate2}, and so forth. It should be noted that these protected layouts are generated to be used along with the ECO flow, and the finalized layout, which includes all potential online checkers and is intended for fabrication, is derived after the conclusion of the ECO round.

Since the difference between each protected layout and its predecessor is the addition of only one online monitor, the user can decide whether to keep or discard the introduced online monitor. This incremental approach ensures a controlled integration of online monitors into the design while precisely managing the resources utilized for adding each online monitor. Moreover, it allows for a fine-grained assessment of the impact on the PPA parameters of the design.

\subsection{ECO Flow}
As described, the insertion of online checkers and the generation of protected layouts are efficiently conducted in our methodology, prioritizing area and resource utilization. We enhance this approach by introducing a timing-aware element to the ECO flow. We introduce a metric called the Degrading Factor (DF), set at 25\% of the total positive slack of the design before the addition of online checkers. This DF serves as a deterministic parameter for deciding whether to keep or discard protected layouts due to the impact on the timing. 

For this purpose, different PPA numbers of the initial layout are stored before incorporating online monitors. The ECO flow starts by selecting the first protected layout, including the highest-ranked online monitor from the cone analysis, and calculates the PPA numbers for this modified layout. Subsequently, the total slack number is compared with the previous layout (the one excluding the newly added monitor). If the slack number is negative or the difference between the new slack and the previous one exceeds the DF, the ECO flow discards the newly added layout since it degrades the timing beyond user constraints and proceeds to the next one. This process continues until all online monitors are integrated, or there is no more slack available for the new logic. Additionally, different checks are performed at each ECO round to ensure compliance with various design rules and avoid issues arising from the application of the new layout.

\section{Experimental Results} \label{sec:results}
This section presents the experimental results of integrating online monitors for five different IPs of OpenTitan. Two IPs were chosen for security reasons: \textsc{alert\_handler\_esc\_timer} has the highest SC, and \textsc{keymgr\_reg\_top} has the lowest SC. Other three IPs were chosen based on their sizes: \textsc{ast\_reg\_top} is the largest design, \textsc{flash\_ctrl\_core\_reg\_top} has an average size, while the smallest one is \textsc{nmi\_gen\_reg\_top}. It should be noted that all these IPs are selected among the ones that already have some assertions, to maintain the concept of repurposing existing assertions. We do not introduce new assertions or modify existing ones across different IPs. The calculated SC for all these IPs is previously illustrated in Fig. \ref{fig:opentitan} and Fig. \ref{fig:sc_ips}.

We use the Cadence suite for all results herein reported: logical synthesis is performed by Genus, while the physical synthesis is performed by Innovus. The formal tool for performing the taint analysis is JasperGold SPV, as mentioned earlier. Our target technology node is a commercial 65nm CMOS one.

\subsection{Impact of Adding Online Monitors on SC}
To assess the impact of integrating online monitors on the security properties of each design, we employ the evaluation scheme outlined in Eq. \ref{eq_sc}. The total covered nodes (\emph{C}) in the numerator of Eq. \ref{eq_sc} now encompass both the nodes previously covered by the assertion and the newly covered nodes introduced by the online monitor. It should be noted that since the output of the assertion or the voter of the online monitors can only be changed by an abnormality in the circuit's expected behavior, the chance of a false positive is eliminated. Consequently, the results obtained from the embedded checkers and assertions are always considered as true positive.

\begin{table*}[ht]
\caption{The impact of adding online monitors on the security of selected IPs} \vspace{-1mm}
\label{table1}
\centering
\begin{tabular}{l|ccccccccc}
IP Name              & Instances       & \begin{tabular}[c]{@{}c@{}}SC \\ Before \end{tabular} & \begin{tabular}[c]{@{}c@{}}SC \\ Total \end{tabular} & \begin{tabular}[c]{@{}c@{}}  SC \\ Added\end{tabular}& \begin{tabular}[c]{@{}c@{}}\# Nodes Covered \\ by  Monitors\end{tabular} & \begin{tabular}[c]{@{}c@{}}\# Applied \\  Monitors\end{tabular} & \begin{tabular}[c]{@{}c@{}}\# Ignored \\  Monitors\end{tabular} & \begin{tabular}[c]{@{}c@{}}\# Total \\  Monitors\end{tabular} & \begin{tabular}[c]{@{}c@{}}Preventing \\ Factor\end{tabular} \\ \hline
\rule{0pt}{8pt}\textsc{\scriptsize alert\_handler\_esc\_timer}  & 1404 & 87.82\%     & 88.25\%    & 0.43\%     & 6                                                                              & 1                                                                   & 0                                                                      & 1                                                                       & Density                                                           \\
\rule{0pt}{8pt}\textsc{\scriptsize ast\_reg\_top}      &7048         & 2.49\%      & 19.94\%    & 17.45\%    & 1382                                                                           & 183                                                                 & 7                                                                    & 190                                                                       & Density                                                         \\
\rule{0pt}{8pt}\textsc{\scriptsize flash\_ctrl\_core\_reg\_top} & 7048 & 1.38\%      & 17.98\%    & 16.6\%     & 954                                                                            & 105                                                                 & 264                                                                    & 369                                                                     & Timing                                                         \\
\rule{0pt}{8pt}\textsc{\scriptsize keymgr\_reg\_top}      &4611      & 0.91\%      & 10.89\%    & 9.98\%     & 490                                                                            & 55                                                                 & 86                                                                     & 141                                                                      & Timing                                                         \\
\rule{0pt}{8pt}\textsc{\scriptsize nmi\_gen\_reg\_top}     & 214     & 3.53\%      & 37.11\%    & 33.58\%    & 92                                                                             & 14                                                                  & 0                                                                     & 14                                                                       & Density                                                          
\end{tabular}
\end{table*}

Table \ref{table1} presents the results concerning the impact of added online monitors on the security of the considered designs. In this table, the first column denotes the IP name, while the second column enumerates the instances in each IP. The third column indicates the SC before the integration of online monitors. These values are obtained by binding all available assertions to each IP and analyzing the coverage using the SPV tool. The fourth and fifth columns represent the SC after adding the online monitors and the increase in SC specifically due to the online monitors, respectively. As indicated, the lowest increase in SC is 0.43\% for the \textsc{alert\_handler\_esc\_timer} IP. This is mainly because this IP already has a high number of covered nodes, and finding suitable candidates that pass all steps of the online monitor insertion flow (as shown in Fig. \ref{fig:backend_flow}) is challenging. In other words, finding a set of uncovered connected nodes with sufficient space around them is more difficult since the total number of uncovered nodes is limited. 
However, the increased SC is not the only benefit we achieve. The introduced logic for inserting online monitors also fills the gaps in the layout and utilizes routing resources, which positively impacts security by reducing potential exploitation opportunities for attackers.

In column 6, the number of total covered nodes is presented after the addition of online monitors. These covered nodes include the nodes not covered by the assertions and the new redundant logic added to form the online checker. Columns 7 and 8 represent the number of applied online monitors in the IP and the number of ignored ones. Column 9 represents the number of online monitors that can be generated for each design after performing density and cone analysis (Fig. \ref{fig:backend_flow}). The total number of online monitors is equal to the number of individual \textit{protected layouts} generated for the ECO flow. It should be noted that not all generated online monitors can be integrated into the design due to timing restrictions. 

The final column indicates the factor preventing the addition of more online monitors. If all available online monitors can be successfully integrated into the design, it means that there are no more online monitors that could be generated, mainly because of the high density around the uncovered nodes. In contrast, if some online monitors are left unembedded in the design (excluding those that exceed the DF), it is mainly because the timing resources of the design are exhausted, and they have to be ignored. More details about the PPA restrictions are discussed in the next part.

\subsection{Impact of Adding Online Monitors on PPA}

The initial layout (baseline) for each IP is configured such that each design has a density ranging between 60\% to 65\%. By doing this, we can preserve a positive setup slack of approximately 10\% of the clock period for each design. Given that the online monitors introduce new logic to the design, influencing its timing, this 10\% margin allows the utilization of positive slack for integrating online monitors.

Table \ref{table2} presents various PPA metrics before and after adding the online monitors to the selected IPs. The first column denotes the IP names. The next two columns illustrate the area and placement characteristics of the layouts, where the second column represents the total area for each design, and the third column represents the placement density. It should be noted that the total area reported in this table refers to the total \textbf{cell area}, and not the overall physical area of the block/chip. Among all designs, \textsc{nmi\_gen\_reg\_top} experienced the most significant increase in area parameters, as it is the smallest design, and even a few online monitors make the size of the added logic comparable to the overall design size.

The fourth column represents the total power consumed by each IP, with \textsc{nmi\_gen\_reg\_top} again showing the largest increase among the IPs due to its small size. The fifth and sixth columns present the timing characteristics of the design. While the hold slack remains relatively constant for all designs, the setup slack undergoes considerable changes for most of the designs, as the redundant logic might impact the timing of the design based on its location in different timing paths of the design. Consistent with Table \ref{table1}, the two designs with the preventing factor of timing (\textsc{flash\_ctrl\_core\_reg\_top} and \textsc{keymgr\_reg\_top}) exhibit the highest decrease in setup slack.

The last column represents the total metal wire length for each design. These metal wires are utilized to connect different parts of the design (signal routing), distribute power, or propagate the clock. As previously mentioned, the adversary not only needs gaps to place malicious logic but also requires available routing resources to connect the malicious logic to other parts of the design. The increase in the total wire length of the design suggests that the design has become more congested, and the free routing resources are now more limited for use by the adversary.

\begin{table*}[htbp]
\caption{The impact of adding online monitors on the PPA metrics of selected IPs} \vspace{-1mm}
\label{table2}
\centering
\begin{tabular}{l|cccccc}
IP Name                             & \begin{tabular}[c]{@{}c@{}}Total Area\\ (\si{\micro\meter\squared})
\end{tabular} & \begin{tabular}[c]{@{}c@{}}Placement\\ Density\end{tabular}  & \begin{tabular}[c]{@{}c@{}}Total Power\\ (mW)\end{tabular} & \begin{tabular}[c]{@{}c@{}}Setup Slack\\ (ns)\end{tabular} & \begin{tabular}[c]{@{}c@{}}Hold Slack\\ (ns)\end{tabular} & \begin{tabular}[c]{@{}c@{}}Total Wire \\Length (\si{\micro\meter})\end{tabular} \\ \hline
\rule{0pt}{8pt}\textsc{\scriptsize alert\_handler\_esc\_timer} (before)           & 3692.88        & 63.94\%             & 1.46        & 0.328                                                         & 0.133                                                          & 22305.40                                                     \\
\rule{0pt}{8pt}\textsc{\scriptsize alert\_handler\_esc\_timer} (after)            & 3700.80    & 64.07\%             & 1.47        & 0.328                                                         & 0.133                                                          & 22351.75                                                     \\

 \rule{0pt}{8pt}Difference           & +0.21\%    & +0.20\%             & +0.68\%        & 0.00\%                                                         & 0.00\%                                                          &  +0.21\% \\

\rowcolor[HTML]{EFEFEF}
\rule{0pt}{8pt}\textsc{\scriptsize ast\_reg\_top} (before)                        & 28644.48   & 61.46\%             & 9.93        & 0.287                                                         & 0.084                                                          & 353799.80                                                    \\

\rowcolor[HTML]{EFEFEF}
\rule{0pt}{8pt}\textsc{\scriptsize ast\_reg\_top} (after)                         & 30848.76   & 66.18\%             & 10.26       & 0.091                                                         & 0.07                                                          & 397612.20                                                    \\

 \rowcolor[HTML]{EFEFEF}\rule{0pt}{8pt}Difference            & +7.69\%    & +7.68\%             & +3.32\%        & -68.29\%                                                         & -16.67\%                                                          &  +12.38\% \\

\rule{0pt}{8pt}\textsc{\scriptsize flash\_ctrl\_core\_reg\_top} (before)         & 14243.40   & 64.25\%             & 4.94        & 0.210                                                         & 0.181                                                          & 148407.90                                                    \\
\rule{0pt}{8pt}\textsc{\scriptsize flash\_ctrl\_core\_reg\_top} (after)           & 15542.28   & 70.11\%             & 5.13        & 0.007                                                         & 0.186                                                          & 170461.90                                                    \\
 \rule{0pt}{8pt}Difference           & +9.12\%   & +9.12\% & +3.85\%                                                               &  -96.67\%                                                        & +2.76\%                                                          &  +14.86\% \\
\rowcolor[HTML]{EFEFEF}
\rule{0pt}{8pt}\textsc{\scriptsize keymgr\_reg\_top} (before)                    & 18325.80   & 62.68\%             & 8.11        & 0.278                                                         & 0.152                                                          & 186978.10                                                    \\
\rowcolor[HTML]{EFEFEF}
\rule{0pt}{8pt}\textsc{\scriptsize keymgr\_reg\_top} (after)                    & 19062.72   & 65.20\%              & 8.29        & 0.000                                                         & 0.151                                                              & 199832.90                                                    \\
 \rowcolor[HTML]{EFEFEF}\rule{0pt}{8pt}Difference        & +4.02\%   & +4.02\%     &+2.22\%    & -100\%                                                                                                               & -0.66\%                                                          &  +6.87\% \\
\rule{0pt}{8pt}\textsc{\scriptsize nmi\_gen\_reg\_top} (before)                   & 769.68     & 61.16\%             & 0.23        & 0.118                                                         & 0.178                                                          & 4841.51                                                      \\
 \rule{0pt}{8pt}\textsc{\scriptsize nmi\_gen\_reg\_top} (after)                     & 918.00     & 72.94\%             & 0.27        & 0.043                                                         & 0.179                                                          & 6077.74   \\                                                  
 
  \rule{0pt}{8pt}Difference           & +19.27\%  & +19.26\%  & +17.39\%        & -63.56\%                                                         & +0.56\%                                                          &  +25.53\% \\
\end{tabular}
\end{table*}

To provide a more detailed insight into the impact of adding each online monitor to the design characteristics, we have extracted various PPA results at the end of each ECO round, where the new online monitor is successfully integrated into the design. This detailed breakdown allows us to observe individual impacts on different characteristics throughout the iterative ECO process. Fig.~\ref{fig:setup_slack} illustrates the degradation in the setup slack after each successful ECO round. The vertical axis represents the total setup slack time in nanoseconds, ranging from the worst to the best slack time, while the horizontal axis depicts the progression of ECO rounds.

As previously mentioned, the ECO flow initiates with the layout containing the assertions, referred to as the ``Baseline" in Fig.~\ref{fig:setup_slack}. As depicted in this figure, some rounds have a negligible impact on the timing, while others considerably degrade the total setup slack. There are also cases in which the slack is improved, attributed to the heuristics of the physical synthesis tool. However, this degradation does not exceed the DF, which is set at 25\% of the total setup slack. It is worth noting that this parameter can be adjusted based on the user's preferences. For instance, if set to lower values, online monitors causing a sudden decrease in the total setup slack (e.g., the online monitor added in round 81 in Fig.~\ref{fig:setup_slack}\subref{setup_ast}) will be discarded, resulting in a smoother overall trend for setup slack decrease.

\definecolor{color21}{RGB}{215,25,28} 
\definecolor{color22}{RGB}{253,174,97} 
\definecolor{color23}{RGB}{8,81,156} 
\definecolor{color24}{RGB}{166,217,106} 
\definecolor{color25}{RGB}{26,150,65} 

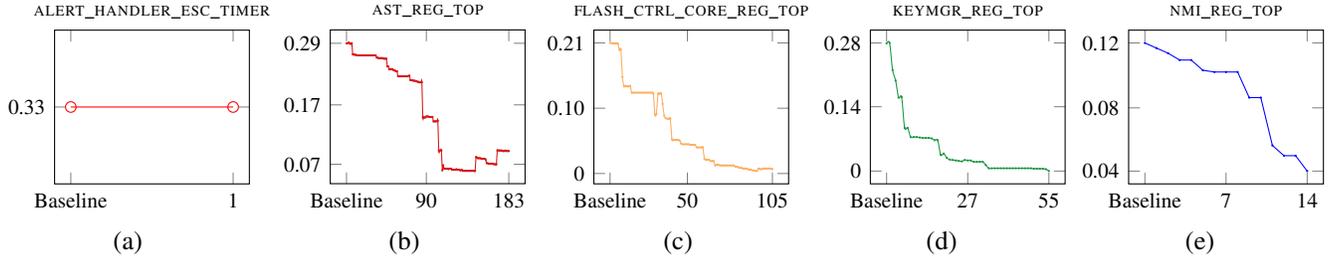
\begin{figure*}[!]
  \centering
  
\subfloat[]{
\begin{tikzpicture}
    \begin{axis}[
      title={\textsc{\scriptsize alert\_handler\_esc\_timer}},
      title style={font=\footnotesize, yshift=-.15cm},
      grid=none,
      width=0.23\textwidth,
      height=0.2\textwidth,
      xtick={0, 1}, 
      xticklabels={Baseline, 1},
      ytick={.328}, 
      yticklabels={0.33},
      cycle list name=color list, 
      tick label style={font=\footnotesize},
    ]
      \addplot+[mark=o, smooth] coordinates {
        (0, 0.328)
        (1, 0.328)
      };
    \end{axis}
  \end{tikzpicture}\hspace{-3mm}
\label{setup_alert}}
  \subfloat[]{
  \begin{tikzpicture}
    \begin{axis}[
      title={\textsc{\scriptsize ast\_reg\_top}},
      title style={font=\footnotesize, yshift=-.13cm},
      grid=none,
      width=0.23\textwidth,
      height=0.20\textwidth,
      xtick={0, 90, 183}, 
      xticklabels={Baseline, 90,  183},
      ytick={0.067, 0.175, 0.287}, 
      yticklabels={0.07, 0.17, 0.29},
      tick label style={font=\footnotesize},
    cycle list name=color list, 
    ]
      \addplot+[mark=o, smooth, color=color21, mark size=0.2] coordinates {
    (0, 0.287) (1, 0.287)   (2, 0.288) (3, 0.288) (4, 0.286) (5, 0.286) (6, 0.287) (7, 0.267) (8, 0.267) (9, 0.266) (10, 0.266) (11, 0.266) (12, 0.265) (13, 0.265) (14, 0.265) (15, 0.265) (16, 0.265) (17, 0.265) (18, 0.265) (19, 0.265) (20, 0.265) (21, 0.265) (22, 0.265) (23, 0.265) (24, 0.265) (25, 0.265) (26, 0.265)  (27, 0.265) (28, 0.265) (29, 0.265) (30, 0.265) (31, 0.265) (32, 0.265) (33, 0.265) (34, 0.261) (35, 0.261) (36, 0.26) (37, 0.26) (38, 0.26) (39, 0.26) (40, 0.26) (41, 0.26) (42, 0.259) (43, 0.259) (44, 0.259) (45, 0.259) (46, 0.245) (47, 0.245) (48, 0.24) (49, 0.24) (50, 0.24) (51, 0.24) (52, 0.24) (53, 0.238) (54, 0.238) (55, 0.237) (56, 0.236) (57, 0.236) (58, 0.227) (59, 0.227) (60, 0.227) (61, 0.227) (62, 0.227) (63, 0.227) (64, 0.227) (65, 0.227) (66, 0.227) (67, 0.227) (68, 0.227) (69, 0.227) (70, 0.228) (71, 0.228) (72, 0.22) (73, 0.22) (74, 0.22)  (75, 0.219) (76, 0.219) (77, 0.219) (78, 0.218) (79, 0.218) (80, 0.217) (81, 0.217) (82, 0.216) (83, 0.216) (84, 0.216) (85, 0.216) (86, 0.153) (87, 0.153) (88, 0.152) (89, 0.152) (90, 0.154) (91, 0.154) (92, 0.154) (93, 0.153) (94, 0.153) (95, 0.153) (96, 0.153) (97, 0.153) (98, 0.145) (99, 0.145) (100, 0.145) (101, 0.145) (102, 0.145) (103, 0.146) (104, 0.092) (105, 0.092) (106, 0.092) (107, 0.092) (108, 0.056) (109, 0.065) (110, 0.059) (111, 0.059) (112, 0.059) (113, 0.059) (114, 0.059) (115, 0.058) (116, 0.059) (117, 0.059) (118, 0.059) (119, 0.057) (120, 0.057) (121, 0.057) (122, 0.057) (123, 0.057) (124, 0.058) (125, 0.057) (126, 0.057) (127, 0.057) (128, 0.056) (129, 0.056) (130, 0.057) (131, 0.055) (132, 0.055) (133, 0.055) (134, 0.055) (135, 0.055) (136, 0.055) (137, 0.055) (138, 0.055) (139, 0.055) (140, 0.055) (141, 0.055) (142, 0.055) (143, 0.055) (144, 0.055) (145, 0.055) (146, 0.08) (147, 0.079) (148, 0.078) (149, 0.078) (150, 0.078) (151, 0.078) (152, 0.077) (153, 0.077) (154, 0.077) (155, 0.077) (156, 0.077) (157, 0.075) (158, 0.069) (159, 0.069) (160, 0.068) (161, 0.068) (162, 0.068) (163, 0.068) (164, 0.068) (165, 0.068) (166, 0.067) (167, 0.067) (168, 0.067) (169, 0.067) (170, 0.092) (171, 0.092) (172, 0.092) (173, 0.092) (174, 0.092) (175, 0.092)  (176, 0.091) (177, 0.091) (178, 0.091) (179, 0.092) (180, 0.091) (181, 0.091) (182, 0.091) (183, 0.091)
      };
    \end{axis}
  \end{tikzpicture}
  \label{setup_ast}}\hspace{-3mm}
\subfloat[]{
\begin{tikzpicture}
    \begin{axis}[
      title={\textsc{\scriptsize flash\_ctrl\_core\_reg\_top}},
      title style={font=\footnotesize, yshift=-.15cm},
      grid=none,
      width=0.23\textwidth,
      height=0.20\textwidth,
      xtick={0, 50, 105}, 
      xticklabels={Baseline, 50, 105},
      ytick={0, 0.105, 0.210}, 
      tick label style={font=\footnotesize},
      yticklabels={0, 0.10, 0.21},
    cycle list name=color list, 
    ]
      \addplot+[mark=o, smooth, color=color22, mark size=0.2] coordinates {
    (0, 0.21) (1, 0.21) (2, 0.209) (3, 0.209) (4, 0.209) (5, 0.209) (6, 0.199) (7, 0.199) (8, 0.155) (9, 0.14) (10, 0.141) (11, 0.14) (12, 0.14) (13, 0.141) (14, 0.13) (15, 0.13) (16, 0.13) (17, 0.13) (18, 0.13) (19, 0.13) (20, 0.13) (21, 0.13) (22, 0.13) (23, 0.13) (24, 0.13) (25, 0.13) (26, 0.13) (27, 0.13) (28, 0.128) (29, 0.095) (30, 0.096) (31, 0.128) (32, 0.128) (33, 0.129)  (34, 0.112) (35, 0.093) (36, 0.089) (37, 0.088) (38, 0.088) (39, 0.088) (40, 0.055) (41, 0.054)  (42, 0.054) (43, 0.054) (44, 0.054)  (45, 0.053) (46, 0.048) (47, 0.047) (48, 0.047) (49, 0.047)  (50, 0.047) (51, 0.047) (52, 0.046)  (53, 0.046) (54, 0.046) (55, 0.046) (56, 0.042) (57, 0.042)  (58, 0.042) (59, 0.042) (60, 0.042)  (61, 0.024) (62, 0.022) (63, 0.022) (64, 0.022) (65, 0.02) (66, 0.02) (67, 0.02) (68, 0.013) (69, 0.014) (70, 0.014) (71, 0.013) (72, 0.013) (73, 0.013) (74, 0.013)  (75, 0.013) (76, 0.013) (77, 0.013) (78, 0.013)  (79, 0.013) (80, 0.013) (81, 0.011) (82, 0.011)  (83, 0.01)  (84, 0.009) (85, 0.009) (86, 0.008) (87, 0.008) (88, 0.008) (89, 0.007) (90, 0.006) (91, 0.006) (92, 0.005)  (93, 0.005) (94, 0.004) (95, 0.004) (96, 0.008) (97, 0.007) (98, 0.007) (99, 0.008) (100, 0.008) (101, 0.008) (102, 0.008) (103, 0.008) (104, 0.008) (105, 0.007)
      };
    \end{axis}
  \end{tikzpicture}
\label{setup_flash}}\hspace{-4mm}
\subfloat[]{
\begin{tikzpicture}
    \begin{axis}[
      title={\textsc{\scriptsize keymgr\_reg\_top}},
      title style={font=\footnotesize, yshift=-.15cm},
      grid=none,
      width=0.23\textwidth,
      height=0.20\textwidth,
      xtick={0, 27, 54}, 
      xticklabels={Baseline, 27, 55},
      ytick={0, 0.14, 0.279}, 
      yticklabels={0, 0.14, 0.28},
      tick label style={font=\footnotesize},
    cycle list name=color list, 
    ]
      \addplot+[mark=o, smooth, color=color25, mark size=0.2] coordinates {
    (0, 0.278) (1, 0.279) (2, 0.22) (3, 0.197) (4, 0.16) (5, 0.161) (6, 0.094) (7, 0.094) (8, 0.074) (9, 0.074) (10, 0.074) (11, 0.073) (12, 0.072) (13, 0.072) (14, 0.072) (15, 0.071) (16, 0.067) (17, 0.067) (18, 0.035) (19, 0.038) (20, 0.029) (21, 0.025) (22, 0.024) (23, 0.023) (24, 0.022) (25, 0.021) (26, 0.024) (27, 0.023) (28, 0.023) (29, 0.02) (30, 0.02) (31, 0.02) (32, 0.02) (33, 0.013) (34, 0.006) (35, 0.006) (36, 0.006) (37, 0.006) (38, 0.006) (39, 0.006) (40, 0.006) (41, 0.006) (42, 0.006) (43, 0.006) (44, 0.006) (45, 0.006) (46, 0.006) (47, 0.006) (48, 0.006) (49, 0.005) (50, 0.005) (51, 0.005) (52, 0.005) (53, 0.004) (54, 0)
      };
    \end{axis}
  \end{tikzpicture}
\label{setup_key}}\hspace{-3mm}
\subfloat[]{
\begin{tikzpicture}
    \begin{axis}[
      title={\textsc{\scriptsize nmi\_reg\_top}},
      title style={font=\footnotesize, yshift=-.15cm},
      grid=none,
      width=0.23\textwidth,
      height=0.20\textwidth,
      xtick={0, 7 ,14}, 
      xticklabels={Baseline, 7, 14},
      ytick={ 0.043, 0.08, 0.118}, 
      yticklabels={ 0.04, 0.08, 0.12},
      tick label style={font=\footnotesize},
    cycle list name=color list, 
    ]
      \addplot+[mark=o, color=blue, mark size=0.2] coordinates {
       (0, 0.118) (1, 0.115) (2, 0.112) (3, 0.108) (4, 0.108) (5, 0.102) (6, 0.101) (7, 0.101) (8, 0.101) (9, 0.086) (10, 0.086) (11, 0.058) (12, 0.052) (13, 0.052) (14, 0.043)
      };
    \end{axis}
  \end{tikzpicture}
\label{setup_nmi}}

  \caption{Changes in setup slack after each round of adding the online monitors for different IPs}
  \label{fig:setup_slack}
\end{figure*}

Fig.~\ref{fig:wire_change} illustrates the progressive increase in wire length for each metal layer in the protected layouts. The vertical axis represents the wire length in \si{\micro\meter}, while the horizontal axis denotes the protected layout in each round. Similar to Fig.~\ref{fig:setup_slack}, the ``Baseline" term refers to the layout containing assertions without online monitors. Although the number of available metal stacks varies in different target technologies, with new technologies typically offering ten or more metal layers, higher consumption of upper metal layers indicates increased congestion in the design. As a defender, our focus is on higher consumption of upper metal layers, reflecting the overall increase in congestion in the protected layouts. Excluding \textsc{alert\_handler\_esc\_timer}, where only one online checker was available to be added, we observe a consistent trend of increased wire length in metal layers M3-M6 for all designs in Fig.~\ref{fig:wire_change}, thereby reducing the available free routing resources for potential adversaries.


\begin{figure*}[!]
  \centering
\subfloat[]{
 \begin{tikzpicture}
    \begin{axis}[
      title={\textsc{\scriptsize alert\_handler\_esc\_timer}},
      title style={font=\scriptsize, yshift=-.25cm},
      grid=none,
      width=0.20\textwidth,
      height=0.21\textwidth,
      xtick={1, 2}, 
      xticklabels={Baseline, 1},
      ytick={0, 9688}, 
      yticklabels={0, 10k},
      axis lines=left,
      cycle list name=color list, 
      tick label style={font=\scriptsize, xshift=1mm},
      enlargelimits=0.05,
      scaled y ticks=false,
    ]
    \addplot[mark=o,cyan, mark size=0.4] coordinates {
    (1, 666.1) (2, 664.29) 
    };

    \addplot[mark=square, color21, mark size=0.4] coordinates {
        (1, 6624.64) (2, 6628.19) 
    };

    \addplot[mark=triangle,color25, mark size=0.4] coordinates {
        (1, 9654.86) (2,9688.86) 
    };

    \addplot[mark=diamond,orange, mark size=0.4] coordinates {
        (1, 3837.8) (2, 3848.4) 
    };

    \addplot[mark=pentagon, magenta, mark size=0.4] coordinates {
        (1, 1503.6) (2, 1503.6) 
    };

    \addplot[mark=x,brown , mark size=0.4] coordinates {
        (1, 18.4) (2, 18.4) 
    };

    \addplot[mark=square*,blue, mark size=0.4] coordinates {
        (1, 0) (2, 0) 
    };

    \end{axis}
  \end{tikzpicture}
} 
\subfloat[]{
  \begin{tikzpicture}
    \begin{axis}[
      title={\textsc{\scriptsize ast\_reg\_top}},
      title style={font=\scriptsize, yshift=-.25cm},
      grid=none,
      width=0.20\textwidth,
      height=0.21\textwidth,
      xtick={1, 90, 184}, 
      xticklabels={Baseline, 90, 183},
      ytick={3500, 104000}, 
      yticklabels={0, 104k},
      axis lines=left,
      cycle list name=color list, 
      tick label style={font=\scriptsize, xshift=1mm},
      enlargelimits=0.05,
      scaled y ticks=false,
    ]
    \addplot[mark=o,cyan, mark size=0.4] coordinates {
    (1, 8441.06) (2, 8438.26) (3, 8421.66) (4, 8420.26) (5, 8418.45) (6, 8414.66) (7, 8414.06) (8, 8415.45) (9, 8410.9) (10, 8408.53) (11, 8408.38) (12, 8407.81) (13, 8405.01) (14, 8400.41) (15, 8384.6) (16, 8381.4) (17, 8382.8) (18, 8383) (19, 8386.4) (20, 8385.93) (21, 8380.34) (22, 8379.34) (23, 8380.22) (24, 8378.9) (25, 8377.9) (26, 8377.65) (27, 8373.65) (28, 8369.25) (29, 8365.84) (30, 8365.67) (31, 8365.08) (32, 8364.06) (33, 8366.85) (34, 8359.26) (35, 8358.4) (36, 8354.2) (37, 8361.8) (38, 8346.43) (39, 8347.03) (40, 8347.23) (41, 8346.18) (42, 8349.38) (43, 8347.81) (44, 8348.21) (45, 8345.81) (46, 8343.61) (47, 8342.61) (48, 8339.81) (49, 8336.41) (50, 8335.41) (51, 8330.21) (52, 8331.36) (53, 8334.36) (54, 8334.16) (55, 8333.56) (56, 8330.16) (57, 8329.76) (58, 8322.99) (59, 8324.39) (60, 8324.19) (61, 8326.92) (62, 8323.33) (63, 8321.12) (64, 8312.95) (65, 8311.35) (66, 8311.56) (67, 8310.85) (68, 8307.49) (69, 8304.33) (70, 8301.93) (71, 8301.93) (72, 8303.88) (73, 8304.08) (74, 8302.08) (75, 8294.48) (76, 8293.2) (77, 8293.41) (78, 8290.81) (79, 8288.01) (80, 8287.41) (81, 8282.03) (82, 8274.64) (83, 8274.43) (84, 8274.03) (85, 8272.98) (86, 8271.6) (87, 8268.81) (88, 8267.41) (89, 8266.01) (90, 8259.01) (91, 8258.6) (92, 8255.2) (93, 8252.81) (94, 8249.81) (95, 8249.62) (96, 8249.42) (97, 8241.62) (98, 8243.02) (99, 8239.83) (100, 8243.83) (101, 8236.62) (102, 8236.83) (103, 8235.3) (104, 8233.36) (105, 8241.17) (106, 8240.22) (107, 8239.06) (108, 8239.17) (109, 8237.97) (110, 8238.17) (111, 8232.77) (112, 8230.97) (113, 8231.34) (114, 8226.17) (115, 8221.57) (116, 8217.57) (117, 8215.01) (118, 8212.18) (119, 8210.67) (120, 8205.67) (121, 8205.07) (122, 8201.5) (123, 8199.5) (124, 8200.7) (125, 8202.5) (126, 8201.9) (127, 8197.7) (128, 8195.3) (129, 8193.9) (130, 8191.3) (131, 8191.7) (132, 8191.7) (133, 8189.9) (134, 8190.1) (135, 8191.3) (136, 8189.93) (137, 8190.33) (138, 8190.42) (139, 8188.42) (140, 8184.82) (141, 8182.22) (142, 8181.22) (143, 8175.22) (144, 8173.42) (145, 8173.77) (146, 8165.63) (147, 8167.56) (148, 8165.56) (149, 8159.15) (150, 8156.95) (151, 8155.99) (152, 8156.8) (153, 8154.8) (154, 8158.6) (155, 8160.2) (156, 8156.65) (157, 8154.27) (158, 8152.07) (159, 8153.27) (160, 8150.67) (161, 8146.67) (162, 8147.07) (163, 8149.27) (164, 8141.27) (165, 8138.84) (166, 8136.44) (167, 8135.24) (168, 8132.53) (169, 8132.53) (170, 8128.93) (171, 8127.53) (172, 8128.33) (173, 8127.13) (174, 8125.93) (175, 8125.93) (176, 8125.36) (177, 8125.96) (178, 8125.56) (179, 8123.36) (180, 8119.36) (181, 8117.96) (182, 8116.76) (183, 8115.16) (184, 8115.96)
    };

    \addplot[mark=square, mark size=0.4, color21] coordinates {
    (1, 69616.76) (2, 69672.35) (3, 69696.31) (4, 69720.41) (5, 69730.79) (6, 69754.69) (7, 69785.53) (8, 69801.43) (9, 69827.26) (10, 69839.89) (11, 69845.31) (12, 69874.93) (13, 69882) (14, 69892.85) (15, 69892.01) (16, 69895.55) (17, 69901.48) (18, 69903.48) (19, 69923.04) (20, 69932.48) (21, 69936.24) (22, 69947.96) (23, 69961.46) (24, 69954.73) (25, 69973.07) (26, 69996.29) (27, 70035.49) (28, 70048.88) (29, 70058.6) (30, 70080.06) (31, 70104.78) (32, 70145.53) (33, 70171.17) (34, 70180.17) (35, 70192.07) (36, 70178.11) (37, 70239.83) (38, 70295.22) (39, 70312.51) (40, 70318.83) (41, 70340.54) (42, 70374.84) (43, 70390.57) (44, 70386.2) (45, 70422.4) (46, 70450.25) (47, 70467.09) (48, 70483.73) (49, 70516.49) (50, 70580.46) (51, 70575.51) (52, 70595.88) (53, 70614.68) (54, 70622.7) (55, 70645.63) (56, 70684.71) (57, 70681.09) (58, 70704.22) (59, 70713.88) (60, 70747.71) (61, 70766.79) (62, 70800.24) (63, 70804.03) (64, 70828.57) (65, 70843.35) (66, 70863.9) (67, 70880.05) (68, 70881.95) (69, 70905.39) (70, 70915.35) (71, 70955.37) (72, 70971.88) (73, 70988.05) (74, 71010.4) (75, 71026.31) (76, 71029.71) (77, 71052.96) (78, 71066.17) (79, 71081.22) (80, 71091.37) (81, 71090.62) (82, 71103.26) (83, 71112.32) (84, 71124.07) (85, 71102.69) (86, 71121.26) (87, 71144.61) (88, 71142.03) (89, 71159.87) (90, 71192.03) (91, 71215.23) (92, 71223.43) (93, 71264.21) (94, 71276.83) (95, 71279.48) (96, 71303.53) (97, 71349.73) (98, 71344.03) (99, 71381.52) (100, 71461.08) (101, 71462.33) (102, 71475.9) (103, 71496.9) (104, 71501.96) (105, 71527.84) (106, 71554.62) (107, 71578.56) (108, 71590.16) (109, 71612.51) (110, 71641.32) (111, 71660.73) (112, 71656.26) (113, 71662.65) (114, 71681.09) (115, 71691.76) (116, 71700.49) (117, 71716.9) (118, 71730.32) (119, 71743.11) (120, 71766.9) (121, 71782.79) (122, 71808.38) (123, 71845.76) (124, 71855.26) (125, 71880.59) (126, 71879.29) (127, 71914.9) (128, 71904.15) (129, 71977.6) (130, 71984.98) (131, 72001.71) (132, 72012.85) (133, 72030.23) (134, 72063.86) (135, 72041.7) (136, 72036.58) (137, 72062.63) (138, 72090.76) (139, 72104.85) (140, 72108.68) (141, 72131.95) (142, 72136.83) (143, 72162.64) (144, 72165.65) (145, 72187.16) (146, 72199.57) (147, 72223.44) (148, 72244.44) (149, 72241.7) (150, 72268.75) (151, 72258.6) (152, 72266.34) (153, 72277.74) (154, 72259.65) (155, 72263.94) (156, 72279.65) (157, 72296.01) (158, 72305.67) (159, 72342.36) (160, 72338.82) (161, 72344.39) (162, 72368.79) (163, 72406.01) (164, 72424.13) (165, 72432.09) (166, 72454.35) (167, 72459.01) (168, 72477.93) (169, 72496.91) (170, 72507.4) (171, 72562.21) (172, 72570.95) (173, 72583.49) (174, 72604.06) (175, 72615.96) (176, 72610.43) (177, 72629.18) (178, 72631.93) (179, 72634.25) (180, 72652.68) (181, 72665.49) (182, 72678.93) (183, 72693.73) (184, 72716.31)
    };

    \addplot[mark=triangle,color25, mark size=0.4] coordinates {
      (1, 91099.55) (2, 91296.75) (3, 91381.15) (4, 91524.1) (5, 91678.29) (6, 91761.49) (7, 91831.49) (8, 91920.7) (9, 92032.49) (10, 92098.49) (11, 92208.49) (12, 92254.49) (13, 92301.69) (14, 92409.69) (15, 92479.78) (16, 92549.57) (17, 92615.1) (18, 92661.31) (19, 92724.71) (20, 92797.9) (21, 92928.51) (22, 92955.31) (23, 93049.2) (24, 93101.8) (25, 93228.6) (26, 93352.8) (27, 93439.8) (28, 93542.2) (29, 93582.2) (30, 93768.4) (31, 93802.8) (32, 93896.8) (33, 94172.8) (34, 94210.4) (35, 94315.46) (36, 94355.46) (37, 94403.26) (38, 94506.86) (39, 94544.46) (40, 94607.36) (41, 94664.16) (42, 94697.96) (43, 94767.16) (44, 94863.56) (45, 94920.51) (46, 94974.15) (47, 95086.15) (48, 95207.35) (49, 95297.9) (50, 95510.45) (51, 95539.75) (52, 95584.29) (53, 95652.1) (54, 95698.9) (55, 95782.1) (56, 95854.1) (57, 95887.29) (58, 95928.7) (59, 96032.7) (60, 96102.4) (61, 96222.2) (62, 96246.79) (63, 96296.4) (64, 96495.6) (65, 96654.6) (66, 96766.6) (67, 96842.79) (68, 96930.79) (69, 96974.2) (70, 96996.2) (71, 97045.79) (72, 97088.12) (73, 97113.78) (74, 97206.98) (75, 97273.78) (76, 97513.23) (77, 97699.23) (78, 97846.88) (79, 97868.48) (80, 97940.53) (81, 98066.48) (82, 98121.07) (83, 98164.88) (84, 98229.43) (85, 98268.68) (86, 98339.07) (87, 98362.87) (88, 98384.47) (89, 98427.37) (90, 98474.71) (91, 98565.71) (92, 98616.32) (93, 98683.12) (94, 98775.26) (95, 98829.71) (96, 98984.71) (97, 99206.11) (98, 99271.31) (99, 99334.51) (100, 99489.71) (101, 99593.21) (102, 99658.61) (103, 99770.35) (104, 99819.53) (105, 99894.28) (106, 99949.68) (107, 100025.31) (108, 100127.25) (109, 100189.85) (110, 100263.05) (111, 100316.85) (112, 100366.2) (113, 100471.14) (114, 100510.54) (115, 100575.94) (116, 100600.74) (117, 100621.9) (118, 100635.9) (119, 100670.49) (120, 100720.29) (121, 100775.69) (122, 100809.69) (123, 100888.29) (124, 100927.49) (125, 101046.49) (126, 101097.44) (127, 101160.04) (128, 101198.46) (129, 101302.85) (130, 101386.85) (131, 101511.81) (132, 101600.95) (133, 101672.15) (134, 101700.35) (135, 101754.55) (136, 101797.35) (137, 101820.35) (138, 101856) (139, 101944) (140, 101971.8) (141, 102045.65) (142, 102086.79) (143, 102146.99) (144, 102186.79) (145, 102228.79) (146, 102277.2) (147, 102312.15) (148, 102365.4) (149, 102411.4) (150, 102445.79) (151, 102458.39) (152, 102500.74) (153, 102527.74) (154, 102553.15) (155, 102570.95) (156, 102603.15) (157, 102668.9) (158, 102696.7) (159, 102723.49) (160, 102744.7) (161, 102803.7) (162, 102901.49) (163, 102996.9) (164, 103033.9) (165, 103107.9) (166, 103153.7) (167, 103224.04) (168, 103250.65) (169, 103326.85) (170, 103386.4) (171, 103412.4) (172, 103419.79) (173, 103526.99) (174, 103567.6) (175, 103602.2) (176, 103620.25) (177, 103678.4) (178, 103707.8) (179, 103726) (180, 103825.85) (181, 103847.85) (182, 103910.85) (183, 103979.15) (184, 104051.3)
    };

    \addplot[mark=diamond,orange, mark size=0.4] coordinates {
        (1, 86617.6) (2, 86706.2) (3, 86747.2) (4, 86848.8) (5, 86901.8) (6, 86944.8) (7, 86954.6) (8, 87095.8) (9, 87209.2) (10, 87350.8) (11, 87386.2) (12, 87544) (13, 87650.2) (14, 87674.6) (15, 87698.2) (16, 87751.8) (17, 87787.2) (18, 87868.4) (19, 87903) (20, 87930) (21, 87977.2) (22, 88004.6) (23, 88061.8) (24, 88119.2) (25, 88162.4) (26, 88211.6) (27, 88223.8) (28, 88263.4) (29, 88342.6) (30, 88562.2) (31, 88611.8) (32, 88648.8) (33, 88695.4) (34, 88705) (35, 88807.2) (36, 88839.2) (37, 88862.4) (38, 88937.8) (39, 89031.4) (40, 89120) (41, 89161.4) (42, 89200.6) (43, 89364.6) (44, 89597.8) (45, 89655.6) (46, 89734.4) (47, 89770) (48, 89841.2) (49, 89868.4) (50, 89913.6) (51, 89954.4) (52, 90044.4) (53, 90074) (54, 90110.2) (55, 90200.4) (56, 90383) (57, 90418.8) (58, 90499.8) (59, 90653) (60, 90708.4) (61, 90784) (62, 90862.2) (63, 90959.8) (64, 91031.8) (65, 91128) (66, 91218.8) (67, 91394.6) (68, 91465.4) (69, 91583.2) (70, 91685) (71, 91870.6) (72, 91951.8) (73, 92098.8) (74, 92213) (75, 92285) (76, 92419.6) (77, 92536.6) (78, 92646.6) (79, 92685.4) (80, 92716) (81, 92843) (82, 92919) (83, 93031.6) (84, 93155.2) (85, 93203.6) (86, 93276.2) (87, 93366) (88, 93412.6) (89, 93457) (90, 93470.6) (91, 93660.6) (92, 93680) (93, 93728.4) (94, 93828.4) (95, 93889.6) (96, 93959.8) (97, 94005.2) (98, 94066.8) (99, 94114.4) (100, 94153) (101, 94185.8) (102, 94235.2) (103, 94296.4) (104, 94316) (105, 94467.4) (106, 94504) (107, 94544.2) (108, 94677.6) (109, 94722.8) (110, 94762.4) (111, 94806) (112, 94856.2) (113, 94976) (114, 94979.4) (115, 95044.4) (116, 95087.2) (117, 95092.2) (118, 95092.2) (119, 95134.4) (120, 95182.4) (121, 95240.8) (122, 95345) (123, 95365) (124, 95414.8) (125, 95445) (126, 95496.4) (127, 95504.6) (128, 95546.6) (129, 95614.2) (130, 95693.6) (131, 95776) (132, 95799.4) (133, 95844.8) (134, 95877.6) (135, 95952.2) (136, 96028.6) (137, 96069.6) (138, 96111.2) (139, 96213.8) (140, 96252.6) (141, 96314.4) (142, 96387.6) (143, 96439.6) (144, 96475.4) (145, 96605.6) (146, 96654.6) (147, 96678.6) (148, 96838) (149, 96916) (150, 96937.4) (151, 96965) (152, 97018.2) (153, 97046.6) (154, 97089.6) (155, 97114.8) (156, 97147.6) (157, 97192.6) (158, 97239.2) (159, 97266.6) (160, 97300.8) (161, 97348.4) (162, 97380.2) (163, 97423.8) (164, 97451.4) (165, 97476.6) (166, 97502.8) (167, 97574.2) (168, 97609.2) (169, 97629.4) (170, 97752.2) (171, 97804.6) (172, 97821) (173, 97857) (174, 97870.8) (175, 97895.4) (176, 97922.2) (177, 97965.2) (178, 98000.6) (179, 98035) (180, 98102.6) (181, 98143.2) (182, 98189.6) (183, 98251.2) (184, 98300.6)
    };

    \addplot[mark=pentagon,magenta, mark size=0.4] coordinates {
        (1, 69999.8) (2, 70000.6) (3, 70024) (4, 70044.6) (5, 70171) (6, 70182.8) (7, 70189.6) (8, 70225.2) (9, 70251.2) (10, 70271.4) (11, 70334.4) (12, 70352.4) (13, 70366.6) (14, 70437.6) (15, 70476) (16, 70587.2) (17, 70661.8) (18, 70695.4) (19, 70780.6) (20, 70847) (21, 70989.8) (22, 70998.2) (23, 71025.8) (24, 71100.2) (25, 71206) (26, 71305.2) (27, 71337.4) (28, 71456.2) (29, 71475) (30, 71605.6) (31, 71614) (32, 71614) (33, 71615.6) (34, 71615.6) (35, 71640.4) (36, 71643.4) (37, 71643.4) (38, 71756.4) (39, 71775) (40, 71779.6) (41, 71779.6) (42, 71817.6) (43, 71828) (44, 71877.8) (45, 71889.2) (46, 71927) (47, 72046.8) (48, 72148.2) (49, 72148.6) (50, 72155.2) (51, 72168.6) (52, 72206.6) (53, 72240.8) (54, 72259) (55, 72267.6) (56, 72360) (57, 72360) (58, 72363.6) (59, 72455) (60, 72466.2) (61, 72586.6) (62, 72618.6) (63, 72624.6) (64, 72678.2) (65, 72787.4) (66, 72954.6) (67, 73055.4) (68, 73154.8) (69, 73168.6) (70, 73190.6) (71, 73194) (72, 73347.2) (73, 73372.2) (74, 73558) (75, 73560) (76, 73603.6) (77, 73707) (78, 73852.8) (79, 73904.2) (80, 74049.6) (81, 74097.4) (82, 74182.2) (83, 74205.8) (84, 74258.4) (85, 74269.4) (86, 74306.8) (87, 74307.8) (88, 74343.6) (89, 74358.2) (90, 74358.2) (91, 74487) (92, 74513.8) (93, 74556.6) (94, 74730) (95, 74750.4) (96, 74805.6) (97, 74808.8) (98, 74820.4) (99, 74861) (100, 74882.2) (101, 74957) (102, 74957.6) (103, 75040.8) (104, 75070.2) (105, 75085) (106, 75128.8) (107, 75228.6) (108, 75353.4) (109, 75533.2) (110, 75697) (111, 75817) (112, 75960.4) (113, 76131) (114, 76131) (115, 76143) (116, 76166.6) (117, 76180.6) (118, 76180.6) (119, 76206.8) (120, 76248) (121, 76269.4) (122, 76313.8) (123, 76362) (124, 76363.6) (125, 76454.8) (126, 76493.8) (127, 76499.4) (128, 76525.4) (129, 76585) (130, 76707.4) (131, 76806.6) (132, 76927) (133, 76969.6) (134, 76984.4) (135, 77032.2) (136, 77059.6) (137, 77116.6) (138, 77162.8) (139, 77213) (140, 77298.8) (141, 77359) (142, 77397) (143, 77415.8) (144, 77415.8) (145, 77429) (146, 77474) (147, 77472.4) (148, 77495.2) (149, 77574.6) (150, 77575) (151, 77610.6) (152, 77801.8) (153, 77821.4) (154, 77870.4) (155, 77891) (156, 77902.2) (157, 77937.2) (158, 77986.6) (159, 77991.4) (160, 78017.8) (161, 78025.4) (162, 78072.8) (163, 78085.8) (164, 78152.8) (165, 78164.8) (166, 78183.8) (167, 78187.2) (168, 78205) (169, 78226.6) (170, 78234.6) (171, 78270.6) (172, 78281.8) (173, 78281.8) (174, 78318.8) (175, 78372.4) (176, 78393.2) (177, 78481.4) (178, 78742) (179, 78759.8) (180, 78760.8) (181, 78816.8) (182, 78865.4) (183, 78888.6) (184, 78889.4)
    };

    \addplot[mark=x,brown, mark size=0.4] coordinates {
     (1, 17870.8) (2, 17871) (3, 17871) (4, 17871.4) (5, 17921.8) (6, 17956.2) (7, 17985.6) (8, 18008.6) (9, 18049) (10, 18096.6) (11, 18127.6) (12, 18128) (13, 18129.4) (14, 18144) (15, 18159.4) (16, 18178.6) (17, 18178.8) (18, 18194.4) (19, 18223.4) (20, 18263.6) (21, 18308.8) (22, 18308.8) (23, 18308.8) (24, 18321.4) (25, 18321.4) (26, 18321.4) (27, 18328.2) (28, 18328.2) (29, 18399.8) (30, 18523.6) (31, 18550.6) (32, 18550.6) (33, 18550.6) (34, 18550.6) (35, 18621) (36, 18621) (37, 18621) (38, 18703.8) (39, 18786.8) (40, 18786.8) (41, 18786.8) (42, 18857.4) (43, 18963.8) (44, 19169.6) (45, 19192) (46, 19228.6) (47, 19228.6) (48, 19249.6) (49, 19249.6) (50, 19249.6) (51, 19249.6) (52, 19320.2) (53, 19320.2) (54, 19326.6) (55, 19362.6) (56, 19362.6) (57, 19362.6) (58, 19501.4) (59, 19551.4) (60, 19590) (61, 19608.2) (62, 19685.6) (63, 19760) (64, 19821.8) (65, 19861) (66, 19861) (67, 19932.8) (68, 19941.8) (69, 19958.4) (70, 19991) (71, 20027.8) (72, 20127.8) (73, 20281.2) (74, 20300) (75, 20334.2) (76, 20457.6) (77, 20600.4) (78, 20642) (79, 20699.4) (80, 20699.8) (81, 20741.2) (82, 20765.6) (83, 20826.2) (84, 20954.4) (85, 20960.8) (86, 21056) (87, 21111.6) (88, 21127.2) (89, 21139.6) (90, 21139.6) (91, 21154) (92, 21154) (93, 21154) (94, 21208.4) (95, 21248.2) (96, 21248.2) (97, 21346.4) (98, 21391) (99, 21422) (100, 21422) (101, 21496.6) (102, 21497) (103, 21497) (104, 21497) (105, 21512.8) (106, 21512.8) (107, 21546) (108, 21596.8) (109, 21604.2) (110, 21637.6) (111, 21663.4) (112, 21728.8) (113, 21728.8) (114, 21728.8) (115, 21728.8) (116, 21764) (117, 21784.6) (118, 21784.6) (119, 21797.2) (120, 21848) (121, 21911.8) (122, 22031.4) (123, 22031.8) (124, 22031.8) (125, 22041.6) (126, 22166.4) (127, 22167.6) (128, 22172.6) (129, 22205.2) (130, 22236) (131, 22239.8) (132, 22268.6) (133, 22308.6) (134, 22323) (135, 22437.8) (136, 22460.2) (137, 22541.2) (138, 22589) (139, 22620.2) (140, 22646.4) (141, 22646.4) (142, 22769.2) (143, 22769.2) (144, 22769.2) (145, 22893.4) (146, 23003.8) (147, 23004.2) (148, 23075.4) (149, 23261.8) (150, 23261.8) (151, 23305.2) (152, 23354.2) (153, 23355) (154, 23394.2) (155, 23434) (156, 23460) (157, 23482.2) (158, 23482.2) (159, 23488) (160, 23539.6) (161, 23572.4) (162, 23573.2) (163, 23574) (164, 23667.8) (165, 23697.4) (166, 23730.8) (167, 23745.6) (168, 23783.2) (169, 23783.6) (170, 23783.6) (171, 23807.2) (172, 23977.2) (173, 23977.2) (174, 24009.6) (175, 24043.8) (176, 24066.2) (177, 24066.2) (178, 24113.8) (179, 24247.8) (180, 24254) (181, 24289.2) (182, 24348.8) (183, 24348.8) (184, 24348.8)
    };

    \addplot[mark=square*,blue, mark size=0.4] coordinates {
       (1, 10154.2) (2, 10154.2) (3, 10154.2) (4, 10154.2) (5, 10154.2) (6, 10154.2) (7, 10154.2) (8, 10155.6) (9, 10155.6) (10, 10155.6) (11, 10155.6) (12, 10155.6) (13, 10155.6) (14, 10155.6) (15, 10155.6) (16, 10155.6) (17, 10183) (18, 10183) (19, 10183) (20, 10183) (21, 10183) (22, 10183) (23, 10183) (24, 10183) (25, 10183) (26, 10183) (27, 10183) (28, 10183) (29, 10183) (30, 10183) (31, 10183) (32, 10183) (33, 10183) (34, 10183) (35, 10204.8) (36, 10204.8) (37, 10204.8) (38, 10204.8) (39, 10204.8) (40, 10204.8) (41, 10204.8) (42, 10204.8) (43, 10204.8) (44, 10204.8) (45, 10204.8) (46, 10204.8) (47, 10204.8) (48, 10204.8) (49, 10204.8) (50, 10204.8) (51, 10204.8) (52, 10204.8) (53, 10204.8) (54, 10204.8) (55, 10204.8) (56, 10204.8) (57, 10204.8) (58, 10204.8) (59, 10204.8) (60, 10204.8) (61, 10206) (62, 10252.6) (63, 10252.6) (64, 10252.6) (65, 10252.6) (66, 10280) (67, 10280) (68, 10280) (69, 10280) (70, 10331.8) (71, 10331.8) (72, 10331.8) (73, 10331.8) (74, 10333) (75, 10333) (76, 10333) (77, 10333) (78, 10333) (79, 10333) (80, 10333) (81, 10353.8) (82, 10371.8) (83, 10371.8) (84, 10386.8) (85, 10386.8) (86, 10386.8) (87, 10386.8) (88, 10386.8) (89, 10386.8) (90, 10386.8) (91, 10386.8) (92, 10387.4) (93, 10387.4) (94, 10387.4) (95, 10387.4) (96, 10387.4) (97, 10387.4) (98, 10387.4) (99, 10387.4) (100, 10387.4) (101, 10387.4) (102, 10387.4) (103, 10387.4) (104, 10387.4) (105, 10387.4) (106, 10441.4) (107, 10464.2) (108, 10464.2) (109, 10464.2) (110, 10577.2) (111, 10605) (112, 10605) (113, 10605) (114, 10605) (115, 10605) (116, 10605) (117, 10605) (118, 10605) (119, 10617.4) (120, 10617.4) (121, 10617.4) (122, 10617.4) (123, 10617.4) (124, 10656.4) (125, 10675.2) (126, 10675.2) (127, 10675.2) (128, 10675.2) (129, 10709.6) (130, 10739.8) (131, 10756.2) (132, 10801.4) (133, 10801.4) (134, 10813.4) (135, 10813.4) (136, 10813.4) (137, 10840.2) (138, 10840.2) (139, 10840.2) (140, 10840.2) (141, 10845.8) (142, 10862.8) (143, 10862.8) (144, 10862.8) (145, 10897.4) (146, 10897.4) (147, 10897.4) (148, 10897.4) (149, 10897.4) (150, 10913.8) (151, 10977.6) (152, 10977.6) (153, 10977.6) (154, 10977.6) (155, 10977.6) (156, 11000.4) (157, 11000.4) (158, 11000.4) (159, 11028.4) (160, 11028.4) (161, 11028.4) (162, 11037.6) (163, 11076) (164, 11088) (165, 11111.8) (166, 11111.8) (167, 11111.8) (168, 11111.8) (169, 11111.8) (170, 11111.8) (171, 11111.8) (172, 11111.8) (173, 11122.6) (174, 11122.6) (175, 11122.6) (176, 11122.6) (177, 11122.6) (178, 11136.8) (179, 11136.8) (180, 11161.8) (181, 11161.8) (182, 11161.8) (183, 11161.8) (184, 11161.8)
    };
    \end{axis}
  \end{tikzpicture}
} 
\subfloat[]{
  \begin{tikzpicture}
    \begin{axis}[
      title={\textsc{\scriptsize flash\_ctrl\_core\_reg\_top}},
      title style={font=\scriptsize, yshift=-.25cm},
      grid=none,
      width=0.20\textwidth,
      height=0.21\textwidth,
      xtick={1, 50, 106}, 
      xticklabels={Baseline, 50, 105},
      ytick={0, 50000}, 
      yticklabels={0, 50k},
      axis lines=left,
      cycle list name=color list, 
      tick label style={font=\scriptsize, xshift=1mm},
      enlargelimits=0.05,
      scaled y ticks=false,
    ]
    \addplot[mark=o,cyan, mark size=0.4] coordinates {
    (1, 3890.12) (2, 3887.32) (3, 3888.32) (4, 3885.12) (5, 3879.72) (6, 3879.14) (7, 3880.24) (8, 3878.47) (9, 3873.51) (10, 3871.7) (11, 3864.93) (12, 3860.93) (13, 3859.74) (14, 3855.14) (15, 3853.34) (16, 3851.74) (17, 3851.45) (18, 3844.79) (19, 3841.99) (20, 3840.59) (21, 3837.59) (22, 3837.59) (23, 3834.28) (24, 3835.28) (25, 3833.28) (26, 3833.28) (27, 3833.48) (28, 3830.88) (29, 3825.34) (30, 3833.49) (31, 3831.68) (32, 3831.49) (33, 3832.09) (34, 3827.45) (35, 3827.89) (36, 3820.72) (37, 3816.72) (38, 3815.93) (39, 3810.53) (40, 3807.33) (41, 3803.12) (42, 3805.53) (43, 3801.53) (44, 3798.72) (45, 3797.12) (46, 3795.72) (47, 3776.72) (48, 3776.72) (49, 3774.72) (50, 3774.72) (51, 3774.88) (52, 3774.47) (53, 3778.28) (54, 3775.68) (55, 3773.08) (56, 3772.47) (57, 3768.28) (58, 3755.08) (59, 3748.68) (60, 3747.11) (61, 3736.31) (62, 3733.85) (63, 3719.45) (64, 3714.45) (65, 3711.65) (66, 3709.51) (67, 3706.68) (68, 3702.71) (69, 3707.22) (70, 3704.22) (71, 3701.82) (72, 3701.87) (73, 3700.66) (74, 3697.66) (75, 3696.87) (76, 3688.16) (77, 3685.36) (78, 3684.66) (79, 3706.12) (80, 3718.72) (81, 3728.76) (82, 3742.27) (83, 3754.67) (84, 3746.07) (85, 3719.67) (86, 3721.91) (87, 3723.16) (88, 3720.26) (89, 3719.66) (90, 3719.51) (91, 3716.93) (92, 3713.7) (93, 3709.7) (94, 3702.49) (95, 3700.89) (96, 3696.49) (97, 3692.33) (98, 3685.13) (99, 3679.93) (100, 3677.53) (101, 3674.38) (102, 3671.98) (103, 3668.98) (104, 3669.41) (105, 3669.41) (106, 3666.41)
    };

    \addplot[mark=square, color21, mark size=0.4] coordinates {
    (1, 35313.64) (2, 35332.69) (3, 35350.71) (4, 35382.62) (5, 35409.69) (6, 35437.33) (7, 35444.44) (8, 35476.71) (9, 35500.01) (10, 35502.25) (11, 35532.45) (12, 35552.53) (13, 35593.79) (14, 35609.51) (15, 35620.83) (16, 35631.43) (17, 35659.11) (18, 35668) (19, 35669.79) (20, 35690.05) (21, 35705.49) (22, 35708.78) (23, 35732.79) (24, 35747.22) (25, 35774.65) (26, 35803.35) (27, 35822.78) (28, 35825.44) (29, 35833.9) (30, 35904.55) (31, 35917.96) (32, 35931) (33, 35951.71) (34, 35965.01) (35, 35962.93) (36, 36009.02) (37, 36026.24) (38, 36120.51) (39, 36137.07) (40, 36152.89) (41, 36172.07) (42, 36227.96) (43, 36251.46) (44, 36264.83) (45, 36284.69) (46, 36299.35) (47, 36328.43) (48, 36358.21) (49, 36371.33) (50, 36425.1) (51, 36424.67) (52, 36439.67) (53, 36486.38) (54, 36503.92) (55, 36506.79) (56, 36495.83) (57, 36483.12) (58, 36475.56) (59, 36486.3) (60, 36497.76) (61, 36530.93) (62, 36519.96) (63, 36542.82) (64, 36534.52) (65, 36562.62) (66, 36582.16) (67, 36601.31) (68, 36634.93) (69, 36681.46) (70, 36694.83) (71, 36702.32) (72, 36718.31) (73, 36744.76) (74, 36757.6) (75, 36795.45) (76, 36806.68) (77, 36807.37) (78, 36811.24) (79, 36810.5) (80, 36834.54) (81, 36847.93) (82, 36846.83) (83, 36859.75) (84, 36872.72) (85, 36865.16) (86, 36882.67) (87, 36930.48) (88, 36937.24) (89, 36934.9) (90, 36952.14) (91, 36961.61) (92, 36967.26) (93, 36978.85) (94, 36966.91) (95, 36987.54) (96, 36999.51) (97, 37009.1) (98, 37018) (99, 37013) (100, 37033) (101, 37025.82) (102, 37043.22) (103, 37051.07) (104, 37063.8) (105, 37067.34) (106, 37082.74)
    };

    \addplot[mark=triangle,color25, mark size=0.4] coordinates {
    (1, 41707.59) (2, 41761.79) (3, 41827.23) (4, 41913.83) (5, 42032.63) (6, 42102.43) (7, 42179.78) (8, 42231.93) (9, 42291.53) (10, 42333.53) (11, 42429.98) (12, 42509.58) (13, 42575.63) (14, 42640.86) (15, 42705.11) (16, 42807.26) (17, 42949.36) (18, 43015.94) (19, 43057.29) (20, 43155.29) (21, 43224.61) (22, 43272.95) (23, 43333.25) (24, 43355.6) (25, 43389.4) (26, 43503.6) (27, 43595.94) (28, 43703.15) (29, 43750.35) (30, 43809.89) (31, 43849.23) (32, 43882.13) (33, 43939.33) (34, 43956.97) (35, 44038.22) (36, 44160.42) (37, 44223.62) (38, 44347.22) (39, 44424.96) (40, 44518.16) (41, 44569.96) (42, 44717.11) (43, 44797.31) (44, 44839.51) (45, 44881.81) (46, 44910.61) (47, 44973.81) (48, 45033.81) (49, 45111.81) (50, 45190.32) (51, 45235.12) (52, 45270.72) (53, 45328.47) (54, 45387.67) (55, 45452.17) (56, 45485.77) (57, 45553.86) (58, 45605.25) (59, 45670.44) (60, 45705.65) (61, 45783.35) (62, 45821.54) (63, 45885.74) (64, 45945.34) (65, 46001.34) (66, 46069.68) (67, 46123.28) (68, 46200.74) (69, 46233.93) (70, 46357.68) (71, 46451.23) (72, 46497.43) (73, 46529.03) (74, 46573.83) (75, 46646.43) (76, 46672.68) (77, 46696.28) (78, 46719.18) (79, 46768.28) (80, 46829.18) (81, 46909.03) (82, 46944.23) (83, 46964.28) (84, 47035.77) (85, 47077.17) (86, 47111.77) (87, 47138.57) (88, 47170.77) (89, 47215.82) (90, 47264.02) (91, 47306.82) (92, 47335.42) (93, 47382.02) (94, 47421.32) (95, 47455.12) (96, 47530.92) (97, 47619.47) (98, 47671.91) (99, 47712.31) (100, 47761.95) (101, 47804.75) (102, 47844.75) (103, 47870.75) (104, 47908.15) (105, 47938.35) (106, 47978.35)
    };

    \addplot[mark=diamond,orange, mark size=0.4] coordinates {
     (1, 43682.6) (2, 43750.2) (3, 43795.8) (4, 43864) (5, 43922.2) (6, 44015.8) (7, 44068.2) (8, 44139.4) (9, 44211.8) (10, 44266.2) (11, 44359.8) (12, 44441.2) (13, 44520.2) (14, 44595.4) (15, 44658.2) (16, 44747.6) (17, 44842) (18, 44936.8) (19, 44990.2) (20, 45077.2) (21, 45133.8) (22, 45165.6) (23, 45193.8) (24, 45249) (25, 45293.6) (26, 45352.2) (27, 45454.2) (28, 45530.2) (29, 45581.4) (30, 45600.8) (31, 45640.8) (32, 45733.2) (33, 45772) (34, 45852.6) (35, 45937.4) (36, 46030) (37, 46091.2) (38, 46211.4) (39, 46351) (40, 46427.2) (41, 46502.4) (42, 46520.8) (43, 46530.8) (44, 46595.2) (45, 46642.6) (46, 46758) (47, 46885.6) (48, 46970) (49, 47062.2) (50, 47202.4) (51, 47239.4) (52, 47292.2) (53, 47317.8) (54, 47415.6) (55, 47453) (56, 47483.2) (57, 47566.6) (58, 47620.2) (59, 47635.4) (60, 47676) (61, 47735.6) (62, 47778.2) (63, 47812.2) (64, 47865.4) (65, 47960.4) (66, 47989) (67, 48011.6) (68, 48117.8) (69, 48231.4) (70, 48309) (71, 48419.4) (72, 48462.6) (73, 48482.6) (74, 48532.8) (75, 48563.4) (76, 48609.4) (77, 48660) (78, 48697.2) (79, 48752.6) (80, 48769.2) (81, 48814.8) (82, 48877.6) (83, 48952) (84, 49001.8) (85, 49070.8) (86, 49096) (87, 49204.2) (88, 49261.4) (89, 49307.6) (90, 49370.6) (91, 49380) (92, 49403) (93, 49486) (94, 49550.4) (95, 49571) (96, 49629.2) (97, 49691) (98, 49702.4) (99, 49748.4) (100, 49870.2) (101, 49897.2) (102, 49922) (103, 49940.4) (104, 49964.2) (105, 50006) (106, 50030)
    };

    \addplot[mark=pentagon,magenta, mark size=0.4] coordinates {
    (1, 21164) (2, 21239.4) (3, 21261) (4, 21313.4) (5, 21367.2) (6, 21415) (7, 21490) (8, 21557.6) (9, 21638.6) (10, 21672.4) (11, 21746.4) (12, 21807.8) (13, 21860.2) (14, 21928.6) (15, 21958.4) (16, 22069.8) (17, 22172.4) (18, 22262.6) (19, 22273.2) (20, 22326.4) (21, 22465) (22, 22473.8) (23, 22474.2) (24, 22502.6) (25, 22604.8) (26, 22659.2) (27, 22729.8) (28, 22831.4) (29, 22865.4) (30, 22866.2) (31, 22866.2) (32, 22866.4) (33, 22912.6) (34, 22932.8) (35, 23006.4) (36, 23047) (37, 23061.8) (38, 23078.4) (39, 23121.2) (40, 23291) (41, 23318) (42, 23376.6) (43, 23376.6) (44, 23386.2) (45, 23425.2) (46, 23464.6) (47, 23468) (48, 23555.2) (49, 23582.8) (50, 23585.8) (51, 23658.8) (52, 23659.2) (53, 23684.2) (54, 23750.8) (55, 23765) (56, 23767.4) (57, 23799.8) (58, 23825.8) (59, 23859.6) (60, 23865) (61, 23876.2) (62, 23897) (63, 23932.8) (64, 23932.8) (65, 23974.4) (66, 23992) (67, 24043.6) (68, 24112.4) (69, 24136.6) (70, 24278.6) (71, 24355.2) (72, 24415) (73, 24420.8) (74, 24448.8) (75, 24450) (76, 24484.4) (77, 24533.8) (78, 24552.2) (79, 24595) (80, 24647.2) (81, 24677.8) (82, 24730.8) (83, 24765.4) (84, 24765.4) (85, 24784.6) (86, 24824.4) (87, 24828.8) (88, 24887.2) (89, 24904) (90, 24908.2) (91, 24908.4) (92, 24955.2) (93, 24967.8) (94, 25003.4) (95, 25039.2) (96, 25237.8) (97, 25279.2) (98, 25284.2) (99, 25284.2) (100, 25455.2) (101, 25515.2) (102, 25518.4) (103, 25542.6) (104, 25543) (105, 25551) (106, 25551)
    };

    \addplot[mark=x,brown, mark size=0.4] coordinates {
    (1, 2446.8) (2, 2446.8) (3, 2446.8) (4, 2461.8) (5, 2486.2) (6, 2507.8) (7, 2533.6) (8, 2533.6) (9, 2623.8) (10, 2668.2) (11, 2705.8) (12, 2767) (13, 2798.6) (14, 2890.8) (15, 2945.4) (16, 2986.4) (17, 3045.2) (18, 3107.8) (19, 3107.8) (20, 3184.8) (21, 3248.2) (22, 3248.2) (23, 3248.2) (24, 3348.2) (25, 3439.8) (26, 3465.4) (27, 3521.4) (28, 3521.4) (29, 3599.6) (30, 3616.8) (31, 3616.8) (32, 3625.6) (33, 3649.6) (34, 3649.6) (35, 3736) (36, 3736) (37, 3770) (38, 3770) (39, 3773.2) (40, 3889.4) (41, 3894.6) (42, 3988.4) (43, 3988.4) (44, 3988.4) (45, 4011.4) (46, 4144.2) (47, 4144.2) (48, 4208) (49, 4310) (50, 4310) (51, 4404.8) (52, 4404.8) (53, 4404.8) (54, 4482) (55, 4494.2) (56, 4494.2) (57, 4550) (58, 4550) (59, 4575) (60, 4575.8) (61, 4663.6) (62, 4663.6) (63, 4663.6) (64, 4663.6) (65, 4688.6) (66, 4707) (67, 4741.8) (68, 4804.2) (69, 4878.2) (70, 4964.2) (71, 5009.2) (72, 5024.8) (73, 5024.8) (74, 5024.8) (75, 5024.8) (76, 5056.2) (77, 5056.2) (78, 5056.2) (79, 5092) (80, 5148.4) (81, 5191.8) (82, 5208) (83, 5261) (84, 5266) (85, 5328.6) (86, 5399.6) (87, 5488.6) (88, 5488.6) (89, 5488.6) (90, 5516.2) (91, 5516.2) (92, 5547.4) (93, 5599.2) (94, 5640.2) (95, 5640.2) (96, 5657.8) (97, 5659.6) (98, 5684.6) (99, 5684.6) (100, 5736.6) (101, 5810) (102, 5810) (103, 5810) (104, 5810) (105, 5825.4) (106, 5825.4)
    };

    \addplot[mark=square*,blue, mark size=0.4] coordinates {
    (1, 103.8) (2, 103.8) (3, 103.8) (4, 103.8) (5, 103.8) (6, 103.8) (7, 103.8) (8, 103.8) (9, 103.8) (10, 103.8) (11, 103.8) (12, 103.8) (13, 103.8) (14, 103.8) (15, 103.8) (16, 103.8) (17, 103.8) (18, 103.8) (19, 103.8) (20, 103.8) (21, 103.8) (22, 103.8) (23, 103.8) (24, 103.8) (25, 103.8) (26, 103.8) (27, 103.8) (28, 103.8) (29, 103.8) (30, 103.8) (31, 103.8) (32, 103.8) (33, 103.8) (34, 103.8) (35, 103.8) (36, 103.8) (37, 103.8) (38, 103.8) (39, 103.8) (40, 103.8) (41, 103.8) (42, 103.8) (43, 103.8) (44, 103.8) (45, 103.8) (46, 103.8) (47, 103.8) (48, 103.8) (49, 103.8) (50, 103.8) (51, 103.8) (52, 103.8) (53, 103.8) (54, 103.8) (55, 103.8) (56, 103.8) (57, 103.8) (58, 103.8) (59, 103.8) (60, 103.8) (61, 103.8) (62, 103.8) (63, 103.8) (64, 103.8) (65, 103.8) (66, 103.8) (67, 103.8) (68, 103.8) (69, 103.8) (70, 103.8) (71, 103.8) (72, 103.8) (73, 103.8) (74, 103.8) (75, 103.8) (76, 103.8) (77, 103.8) (78, 103.8) (79, 103.8) (80, 103.8) (81, 103.8) (82, 103.8) (83, 103.8) (84, 103.8) (85, 103.8) (86, 103.8) (87, 103.8) (88, 103.8) (89, 103.8) (90, 103.8) (91, 103.8) (92, 103.8) (93, 103.8) (94, 103.8) (95, 103.8) (96, 103.8) (97, 103.8) (98, 103.8) (99, 103.8) (100, 103.8) (101, 103.8) (102, 103.8) (103, 103.8) (104, 103.8) (105, 103.8) (106, 103.8)
    };
    \end{axis}
  \end{tikzpicture}
} 
\subfloat[]{
  \begin{tikzpicture}
    \begin{axis}[
      title={\textsc{\scriptsize keymgr\_reg\_top}},
      title style={font=\scriptsize, yshift=-.25cm},
      grid=none,
      legend style={at={(1.2,.5)},anchor=center, font=\footnotesize},
      width=0.20\textwidth,
      height=0.21\textwidth,
      xtick={1, 28, 56}, 
      xticklabels={Baseline, 27, 55},
      ytick={0, 61541}, 
      yticklabels={0, 62k},
      axis lines=left,
      cycle list name=color list, 
      tick label style={font=\scriptsize, xshift=1mm},
      enlargelimits=0.05,
      scaled y ticks=false,
    ]
    \addplot[mark=o,cyan, , mark size=0.4] coordinates {
    (1, 3466.91) (2, 3467.14) (3, 3466.31) (4, 3459.14) (5, 3455.93) (6, 3450.29) (7, 3445.6) (8, 3441.8) (9, 3437.4) (10, 3432.6) (11, 3429.4) (12, 3416.49) (13, 3420.09) (14, 3432.69) (15, 3436.89) (16, 3435.69) (17, 3433.29) (18, 3424.89) (19, 3422.34) (20, 3418.49) (21, 3419.09) (22, 3415.89) (23, 3399.28) (24, 3400.89) (25, 3398.09) (26, 3393.86) (27, 3389.66) (28, 3390.45) (29, 3386.45) (30, 3378.65) (31, 3377.25) (32, 3370.25) (33, 3359.45) (34, 3354.05) (35, 3345.05) (36, 3345.24) (37, 3340.64) (38, 3338.85) (39, 3331.05) (40, 3321.97) (41, 3316.38) (42, 3313.58) (43, 3308.78) (44, 3303.78) (45, 3299.58) (46, 3298.78) (47, 3295.38) (48, 3288.81) (49, 3288.81) (50, 3283.03) (51, 3281.88) (52, 3279.88) (53, 3282.88) (54, 3284.47) (55, 3284.28) (56, 3278.08)
    };

    \addplot[mark=square, color21, mark size=0.4] coordinates {
    (1, 34606.82) (2, 34608.38) (3, 34614.04) (4, 34611.32) (5, 34613.92) (6, 34641.24) (7, 34651.82) (8, 34670.88) (9, 34729.24) (10, 34755.76) (11, 34776.78) (12, 34822.78) (13, 34848.32) (14, 34878.72) (15, 34882.31) (16, 34907.11) (17, 34915.22) (18, 34938.68) (19, 34947.09) (20, 34959.68) (21, 34987.9) (22, 35018.63) (23, 35048.69) (24, 35069.76) (25, 35085.12) (26, 35106.64) (27, 35146.22) (28, 35162.33) (29, 35177.49) (30, 35181.83) (31, 35199.87) (32, 35211.15) (33, 35213.02) (34, 35226.33) (35, 35266.02) (36, 35283.81) (37, 35285.48) (38, 35298.6) (39, 35318.87) (40, 35349.53) (41, 35377.29) (42, 35390.87) (43, 35417.1) (44, 35432.81) (45, 35433.7) (46, 35461.35) (47, 35466.63) (48, 35493.06) (49, 35492.32) (50, 35525.91) (51, 35541.4) (52, 35550.64) (53, 35562.53) (54, 35576.75) (55, 35584.93) (56, 35635.39)
    };

    \addplot[mark=triangle,color25, mark size=0.4] coordinates {
    (1, 57008.4) (2, 57206.15) (3, 57315.15) (4, 57380.75) (5, 57413.35) (6, 57504.58) (7, 57661.18) (8, 57749.32) (9, 57877.72) (10, 57970.32) (11, 58099.32) (12, 58202.32) (13, 58209.51) (14, 58227.92) (15, 58248.11) (16, 58306.32) (17, 58347.86) (18, 58470.18) (19, 58603.38) (20, 58644.97) (21, 58819.58) (22, 58891.36) (23, 59069.57) (24, 59132.76) (25, 59191.17) (26, 59328.07) (27, 59420.96) (28, 59482.56) (29, 59561.01) (30, 59682.01) (31, 59748.81) (32, 59814.61) (33, 59907.21) (34, 59977.36) (35, 60138.71) (36, 60173.71) (37, 60283.11) (38, 60367.9) (39, 60449.51) (40, 60510.11) (41, 60605.51) (42, 60678.31) (43, 60788.26) (44, 60859.65) (45, 60968.26) (46, 61040.86) (47, 61077.06) (48, 61129.86) (49, 61145.46) (50, 61241.26) (51, 61293.61) (52, 61320.21) (53, 61366.81) (54, 61398.01) (55, 61437.36) (56, 61541.65)
    };

    \addplot[mark=diamond,orange, mark size=0.4] coordinates {
    (1, 36818.2) (2, 36874.8) (3, 36891.6) (4, 36926) (5, 36949) (6, 37002.8) (7, 37236.4) (8, 37302.2) (9, 37356.8) (10, 37460.6) (11, 37512.4) (12, 37545.2) (13, 37554) (14, 37554) (15, 37586.6) (16, 37615.6) (17, 37652.8) (18, 37751) (19, 37840) (20, 37957.2) (21, 38050.4) (22, 38093.8) (23, 38150) (24, 38200.2) (25, 38235.6) (26, 38317.4) (27, 38396.8) (28, 38457.2) (29, 38517.8) (30, 38646.4) (31, 38725.2) (32, 38802.4) (33, 38880.6) (34, 38937.8) (35, 38966.8) (36, 39017.2) (37, 39046) (38, 39086) (39, 39142) (40, 39180.6) (41, 39242.4) (42, 39331) (43, 39400.8) (44, 39459.6) (45, 39519.6) (46, 39586.2) (47, 39591.4) (48, 39598.4) (49, 39625.8) (50, 39720.6) (51, 39763.6) (52, 39827.2) (53, 39914.4) (54, 39934) (55, 39989.8) (56, 40069.6)
    };

    \addplot[mark=pentagon,magenta, mark size=0.4] coordinates {
     (1, 49428.6) (2, 49446.4) (3, 49474.2) (4, 49567.8) (5, 49567.8) (6, 49569.4) (7, 49675.4) (8, 49687.4) (9, 49748.2) (10, 49893) (11, 49911.2) (12, 49986) (13, 49986.4) (14, 49986.4) (15, 49986.4) (16, 49986.4) (17, 49986.4) (18, 50058.8) (19, 50127) (20, 50189.4) (21, 50278.8) (22, 50387) (23, 50481.2) (24, 50499.6) (25, 50508.8) (26, 50613) (27, 50726.4) (28, 50744.4) (29, 50780) (30, 50810.4) (31, 50847.2) (32, 50913) (33, 50982.4) (34, 51024) (35, 51060.4) (36, 51101.2) (37, 51176.2) (38, 51184) (39, 51256.8) (40, 51293.8) (41, 51374.2) (42, 51490.2) (43, 51583.4) (44, 51705.4) (45, 51754.6) (46, 51837.2) (47, 51837.2) (48, 51837.2) (49, 51837.2) (50, 51904.8) (51, 51910.6) (52, 51984) (53, 52018) (54, 52048) (55, 52054.4) (56, 52144.6)
    };

    \addplot[mark=x,brown, mark size=0.4] coordinates {
    (1, 1968.8) (2, 1968.8) (3, 1972.2) (4, 1972.2) (5, 1972.2) (6, 1972.2) (7, 2042) (8, 2042) (9, 2070.2) (10, 2180.4) (11, 2202.2) (12, 2214.6) (13, 2214.6) (14, 2214.6) (15, 2214.6) (16, 2214.6) (17, 2214.6) (18, 2248.6) (19, 2284.4) (20, 2315.8) (21, 2360.8) (22, 2404.8) (23, 2404.8) (24, 2426) (25, 2445.2) (26, 2471.8) (27, 2504) (28, 2504) (29, 2511.8) (30, 2585) (31, 2589.6) (32, 2589.6) (33, 2658.2) (34, 2675) (35, 2675) (36, 2675) (37, 2692) (38, 2708.2) (39, 2740) (40, 2755.8) (41, 2754.2) (42, 2754) (43, 2754) (44, 2767) (45, 2775) (46, 2776.2) (47, 2776.2) (48, 2776.2) (49, 2776.2) (50, 2816.8) (51, 2838.4) (52, 2846.6) (53, 2846.6) (54, 2846.6) (55, 2846.6) (56, 2912.4)
    };

    \addplot[mark=square*,blue, mark size=0.4] coordinates {
    (1, 3680.4) (2, 3680.4) (3, 3680.4) (4, 3680.4) (5, 3680.4) (6, 3680.4) (7, 3680.4) (8, 3680.4) (9, 3680.4) (10, 3680.4) (11, 3772.6) (12, 3772.6) (13, 3772.6) (14, 3772.6) (15, 3772.6) (16, 3772.6) (17, 3772.6) (18, 3772.6) (19, 3772.6) (20, 3772.6) (21, 3772.6) (22, 3772.6) (23, 3772.6) (24, 3772.6) (25, 3772.6) (26, 3772.6) (27, 3772.6) (28, 3772.6) (29, 3772.6) (30, 3806.4) (31, 3806.4) (32, 3806.4) (33, 3806.4) (34, 3830.2) (35, 3876.8) (36, 3876.8) (37, 3876.8) (38, 4031.8) (39, 4031.8) (40, 4031.8) (41, 4031.8) (42, 4038.2) (43, 4038.2) (44, 4038.2) (45, 4038.2) (46, 4125) (47, 4139.4) (48, 4139.4) (49, 4139.4) (50, 4139.4) (51, 4139.4) (52, 4139.4) (53, 4251.2) (54, 4251.2) (55, 4251.2) (56, 4251.2) 
    };


    \end{axis}
  \end{tikzpicture}
} 
  \subfloat[]{
\begin{tikzpicture}
    \begin{axis}[
      title={\textsc{\scriptsize nmi\_reg\_top}},
      title style={font=\scriptsize, yshift=-.25cm},
      grid=none,
      legend style={at={(1.3,.5)},anchor=center, font=\tiny},
      width=0.20\textwidth,
      height=0.21\textwidth,
      xtick={1, 15}, 
      xticklabels={Baseline, 14},
      ytick={0, 2794}, 
      yticklabels={0,  2.8k},
      axis lines=left,
      cycle list name=color list, 
      tick label style={font=\scriptsize, xshift=1mm},
      enlargelimits=0.05,
      scaled y ticks=false,
    ]
    \addplot[mark=o,cyan, mark size=0.4] coordinates {
    (1, 175.6) (2, 178.51) (3, 172.51) (4, 169.31) (5, 167.91) (6, 173.11) (7, 169.02) (8, 167.62) (9, 163.53) (10, 161.13) (11, 161.53) (12, 157.73) (13, 157.73) (14, 157.13) (15, 156.53)
    };
    \addlegendentry{M1}

    \addplot[mark=square, color21, mark size=0.4] coordinates {
        (1, 1594.71) (2, 1596.25) (3, 1620.25) (4, 1632.49) (5, 1645.51) (6, 1662.66) (7, 1646.42) (8, 1645.39) (9, 1655.82) (10, 1667.66) (11, 1676.47) (12, 1668.83) (13, 1667.02) (14, 1669) (15, 1672.05)
    };
    \addlegendentry{M2}

    \addplot[mark=triangle,color25, mark size=0.4] coordinates {
        (1, 2305.6) (2, 2412.1) (3, 2438.3) (4, 2461.5) (5, 2489.37) (6, 2540.97) (7, 2610.41) (8, 2624.41) (9, 2646.57) (10, 2668.76) (11, 2689.97) (12, 2720.16) (13, 2767.57) (14, 2785.37) (15, 2794.97)
    };
    \addlegendentry{M3}

    \addplot[mark=diamond,orange, mark size=0.4] coordinates {
        (1, 684.2) (2, 773.2) (3, 809.6) (4, 820.6) (5, 855) (6, 899.2) (7, 945.2) (8, 973.4) (9, 1002.6) (10, 1007) (11, 1071.2) (12, 1080.8) (13, 1225.2) (14, 1284.6) (15, 1304)
    };
    \addlegendentry{M4}

    \addplot[mark=pentagon, magenta, mark size=0.4] coordinates {
        (1, 81.4) (2, 81.4) (3, 81.4) (4, 81.4) (5, 81.4) (6, 81.4) (7, 81.4) (8, 82.2) (9, 110.8) (10, 110.8) (11, 110.8) (12, 117.6) (13, 126.4) (14, 129.6) (15, 137.8)
    };
    \addlegendentry{M5}

    \addplot[mark=x,brown , mark size=0.4] coordinates {
        (1, 0) (2, 0) (3, 0) (4, 0) (5, 0) (6, 0) (7, 0) (8, 12) (9, 12.4) (10, 12.4) (11, 12.4) (12, 12.4) (13, 12.4) (14, 12.4) (15, 12.4)
    };
    \addlegendentry{M6}

    \addplot[mark=square*,blue, mark size=0.4] coordinates {
        (1, 0) (2, 0) (3, 0) (4, 0) (5, 0) (6, 0) (7, 0) (8, 0) (9, 0) (10, 0) (11, 0) (12, 0) (13, 0) (14, 0) (15, 0)
    };
    \addlegendentry{M7}


      
    \end{axis}
  \end{tikzpicture}
}
  \caption{The impact of adding online monitors on the length of different metal layers in the protected layouts for different IPs}
  \label{fig:wire_change}
\end{figure*}
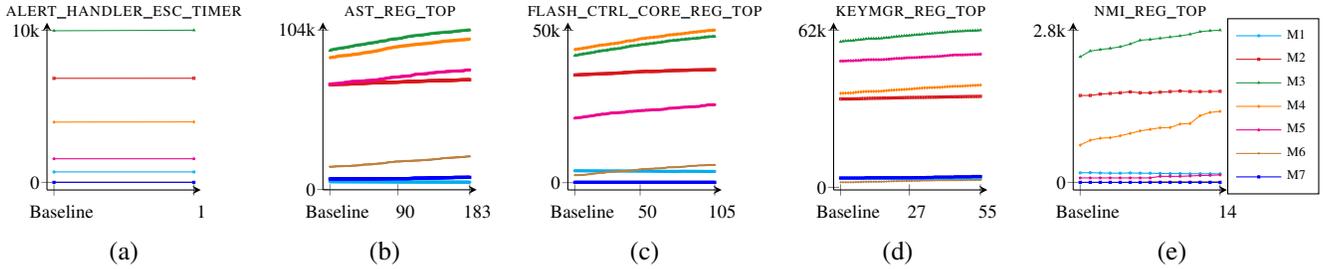


The visual representations of the layout, including placement configuration and the routed view of designs before and after the integration of online monitors, are illustrated in Fig.~\ref{fig:snaps}. In each row, the left image pair showcases the cell placement, while the right one displays the routed view. Specifically, the left image in each pair represents the layout before the addition of online monitors, while the right image corresponds to the final protected layout after the successful completion of all ECO rounds. Upon comparing the images on the right with those on the left, it is evident that the overall placement and routing configuration of the layouts remained unchanged, even for larger designs. This highlights the more efficient utilization of resources, which is a key advantage of our methodology in adding online monitors during the physical synthesis compared to similar works performed in the front-end step of chip design.

\begin{figure*}[!]
    \centering
    \begin{tikzpicture}
        \draw [<->](0,0) -- (0,3.8) node[midway, left, font=\footnotesize] {\: 95.6 \si{\micro\meter}};
    \end{tikzpicture}
      \subfloat[]{\includegraphics[width=0.44\columnwidth, height=0.44\columnwidth]{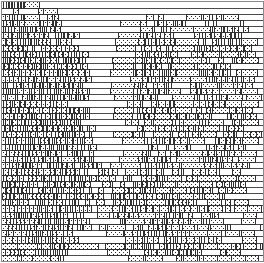}
    \includegraphics[width=0.44\columnwidth, height=0.44\columnwidth]{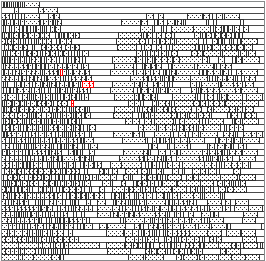} \label{12a}} \hspace{0.005\textwidth} \vspace{-2mm}
    \subfloat[]{\includegraphics[width=0.44\columnwidth, height=0.44\columnwidth]{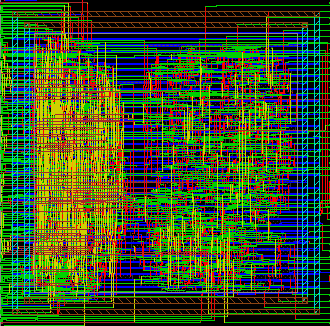}
    \includegraphics[width=0.44\columnwidth, height=0.44\columnwidth]{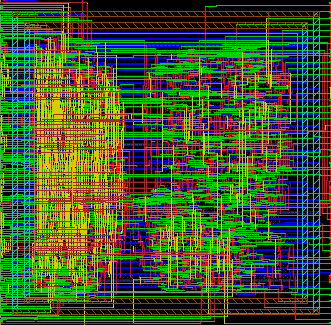} \label{12b}} \hspace{0.05\textwidth} \vspace{-2mm}
   \begin{tikzpicture}
        \draw [<->](0,0) -- (0,3.8) node[midway, left, font=\footnotesize] {234.2 \si{\micro\meter}};
    \end{tikzpicture}
    \subfloat[]{\includegraphics[width=0.44\columnwidth, height=0.44\columnwidth]{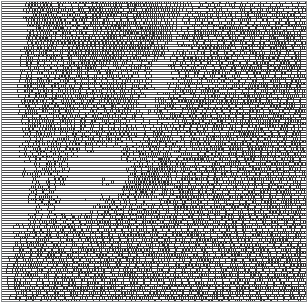}
    \includegraphics[width=0.44\columnwidth, height=0.44\columnwidth]{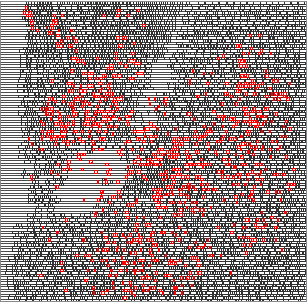} \label{12c}} \hspace{0.005\textwidth}
    \subfloat[]{\includegraphics[width=0.44\columnwidth, height=0.44\columnwidth]{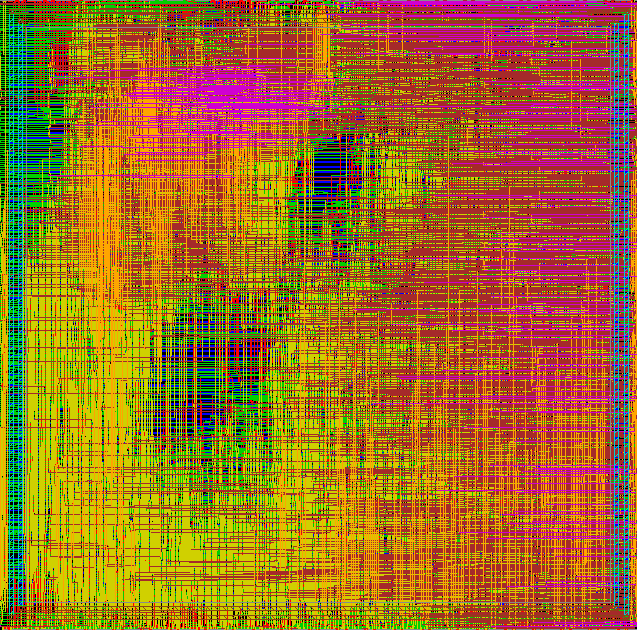}
    \includegraphics[width=0.44\columnwidth, height=0.44\columnwidth]{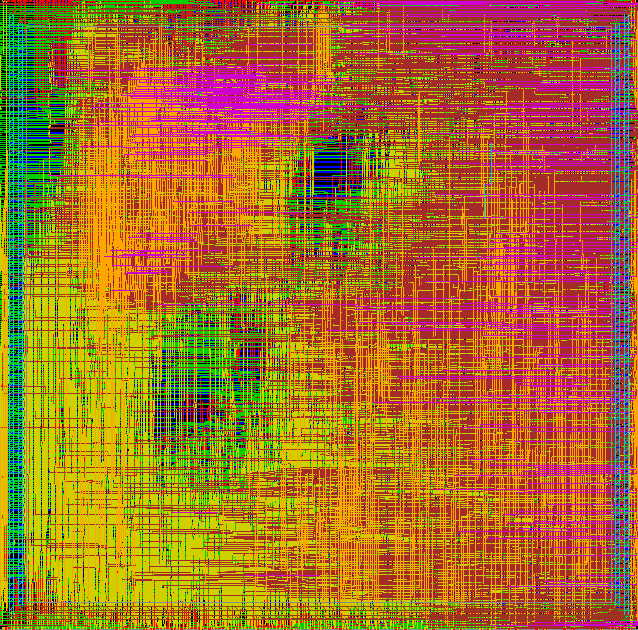} \label{12d}}  \hspace{0.05\textwidth} \vspace{-2mm}
    \begin{tikzpicture}
        \draw [<->](0,0) -- (0,3.8) node[midway, left, font=\footnotesize] {167.6 \si{\micro\meter}};
    \end{tikzpicture}
    \subfloat[]{\includegraphics[width=0.44\columnwidth, height=0.44\columnwidth]{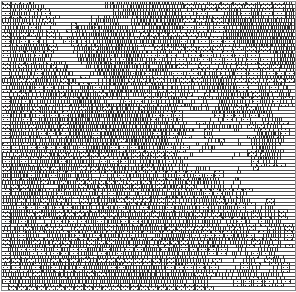}
    \includegraphics[width=0.44\columnwidth, height=0.44\columnwidth]{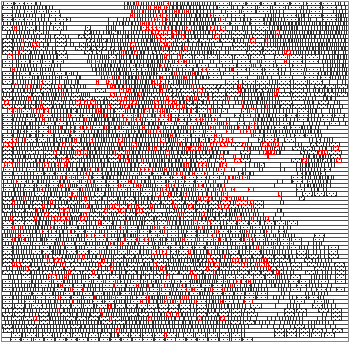} \label{12e}} \hspace{0.005\textwidth}
     \subfloat[]{\includegraphics[width=0.44\columnwidth, height=0.44\columnwidth]{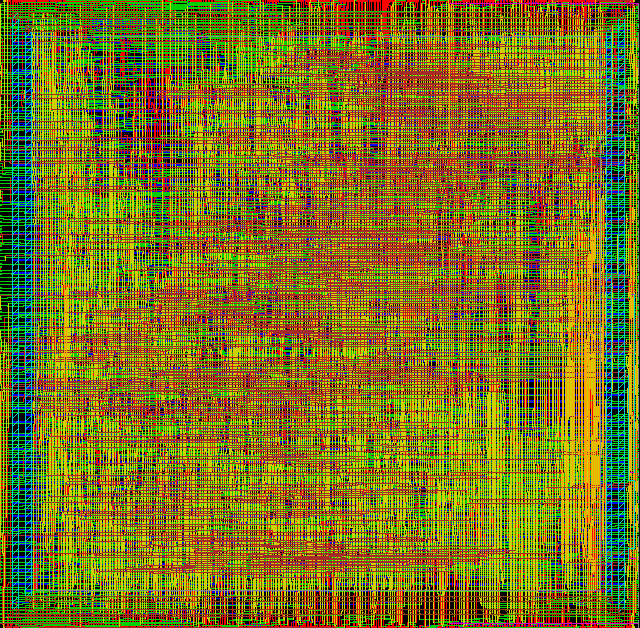}
    \includegraphics[width=0.44\columnwidth, height=0.44\columnwidth]{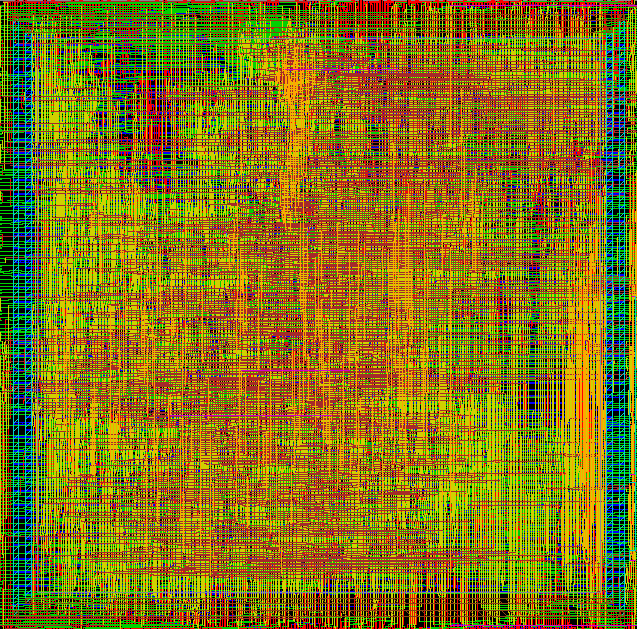} \label{12f}}  \hspace{0.05\textwidth} \vspace{-2mm}
    \begin{tikzpicture}
        \draw [<->](0,0) -- (0,3.8) node[midway, left, font=\footnotesize] {189.2 \si{\micro\meter}};
    \end{tikzpicture}
    \subfloat[]{\includegraphics[width=0.44\columnwidth, height=0.44\columnwidth]{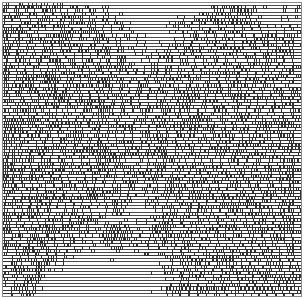}
    \includegraphics[width=0.44\columnwidth, height=0.44\columnwidth]{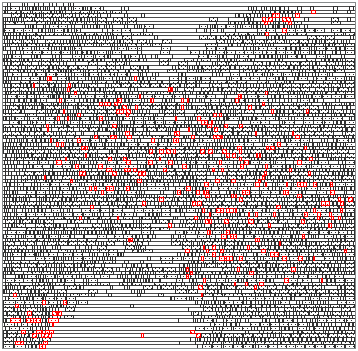} \label{12g}} \hspace{0.005\textwidth}
     \subfloat[]{\includegraphics[width=0.44\columnwidth, height=0.44\columnwidth]{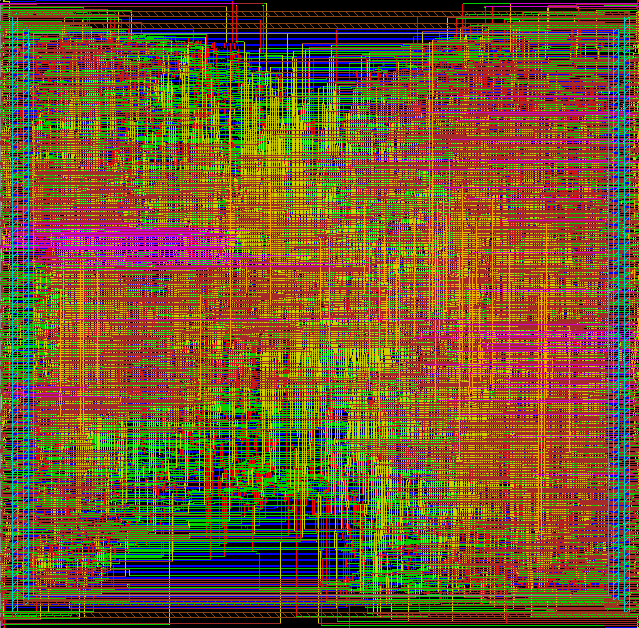}
    \includegraphics[width=0.44\columnwidth, height=0.44\columnwidth]{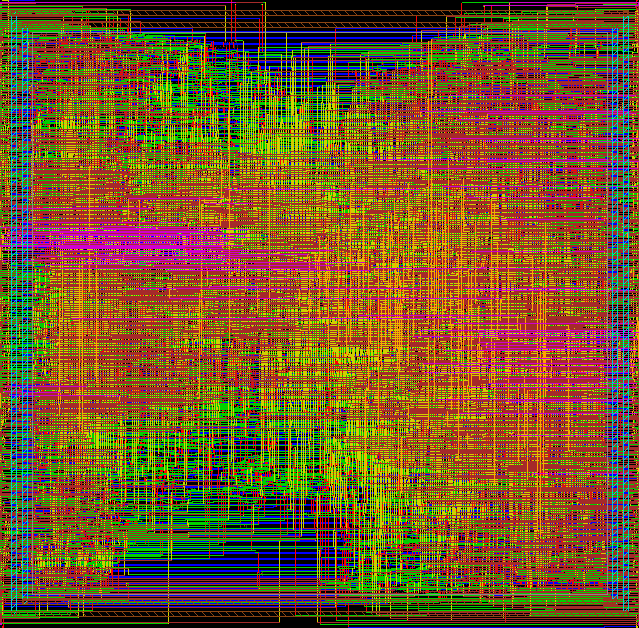} \label{12h}}  \hspace{0.05\textwidth} \vspace{-2mm}
    \begin{tikzpicture}
        \draw [<->](0,0) -- (0,3.8) node[midway, left, font=\footnotesize] {\:46.2 \si{\micro\meter}};
    \end{tikzpicture}
    \subfloat[]{\includegraphics[width=0.44\columnwidth, height=0.44\columnwidth]{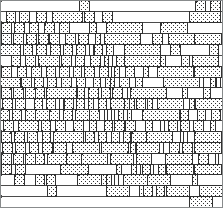}
    \includegraphics[width=0.44\columnwidth, height=0.44\columnwidth]{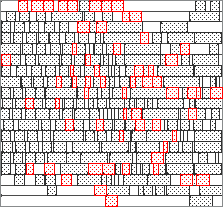} \label{12i}} \hspace{0.005\textwidth}
     \subfloat[]{\includegraphics[width=0.44\columnwidth, height=0.44\columnwidth]{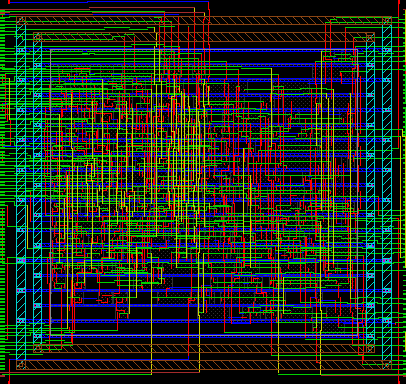}
    \includegraphics[width=0.44\columnwidth, height=0.44\columnwidth]{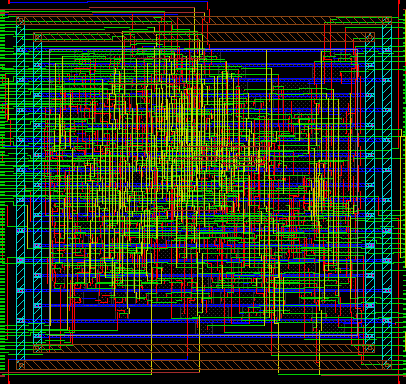} \label{12j}}  \hspace{0.05\textwidth} \vspace{-1mm}
    \caption{The layout view of the IPs before and after adding online monitors, whereas the left pair of images in each row represents the placement configuration, while the right one represents the routed view of each design: a, b) \textsc{\scriptsize alert\_handler\_esc\_timer}, c, d) \textsc{\scriptsize ast\_reg\_top}, e, f) \textsc{\scriptsize flash\_ctrl\_core\_reg\_top}, g, h) \textsc{\scriptsize keymgr\_reg\_top}, and i, j) \textsc{\scriptsize nmi\_reg\_top}}
    \label{fig:snaps}
\end{figure*}
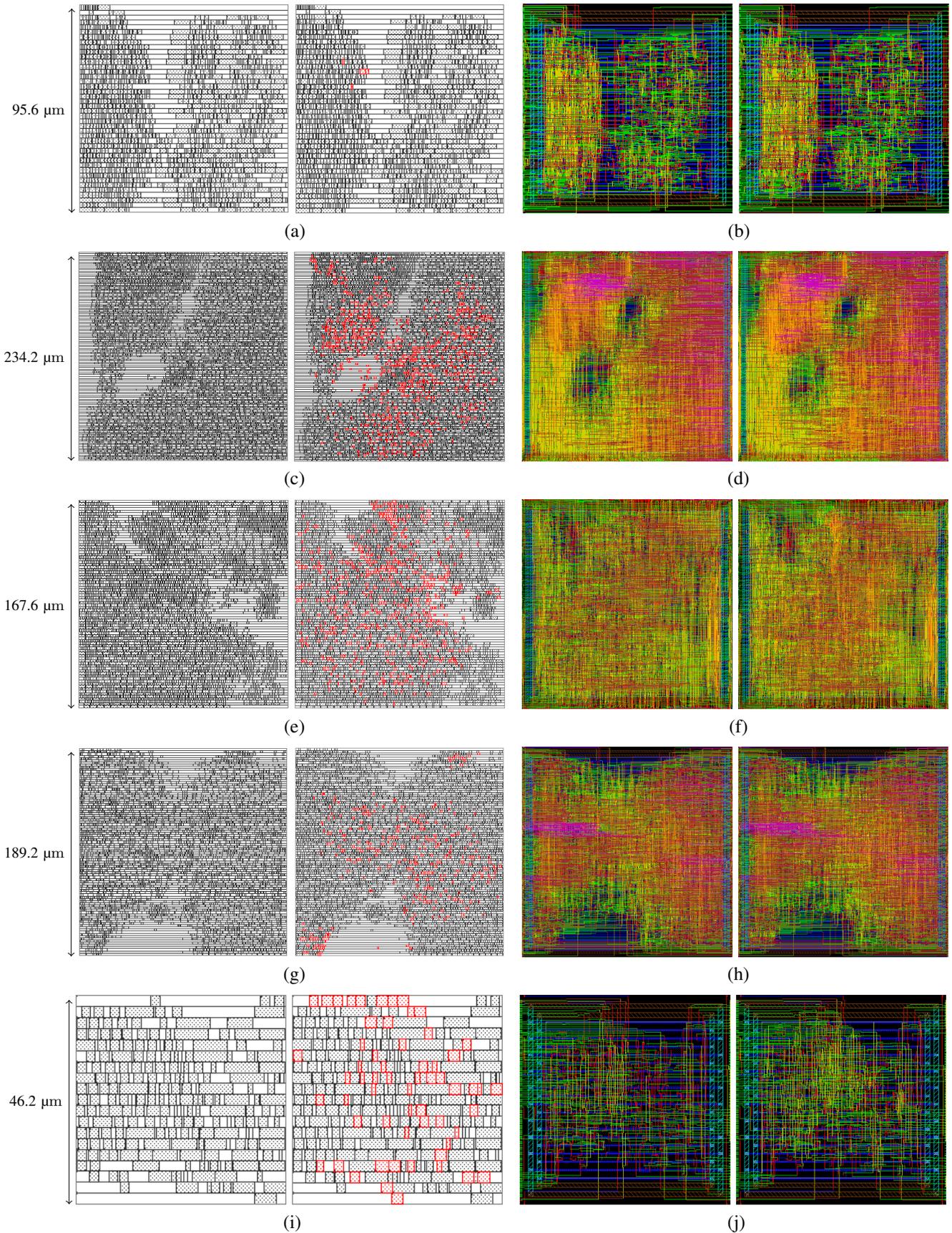

\subsection{SCARF Versus Other Techniques}
Table \ref{tab3} provides a comparative analysis of our framework with other detection and DfHT approaches. The first column indicates the specific technique or category, while the second column references works in that category. The third column categorizes the technique into detection or DfHT (or both). The subsequent column details the chip design stage where the method is applied for protection. Column 5 specifies the location of the potential attacker, and column 6 indicates if the technique can also be employed to prevent HTs. The last two columns briefly outline the advantages and disadvantages of the techniques, respectively.

\begin{table*}[htbp]
\begin{threeparttable}
\caption{Comparison of Different Detection and DfHT Techniques}\label{tab3}
\begin{tabular}{l|ccccccc}
Technique                                                                     & Refs       & \begin{tabular}[c]{@{}c@{}}Detection \\ or DfHT\end{tabular} &  \begin{tabular}[c]{@{}c@{}}Design \\ Stage\end{tabular}        & \begin{tabular}[c]{@{}c@{}}Attacker \\ Location\end{tabular}   & \begin{tabular}[c]{@{}c@{}}Prevents \\ HT?\end{tabular} & Pros.                                                                                                                                               & Cons.                                                                                                                                                                                               \\ \hline

\begin{tabular}[c]{@{}l@{}}IC Fingerprinting,\\ Delay Meas.\end{tabular} &      \cite{14,15,16,17}      & Detection                                                                                                                & Test                                                           & \begin{tabular}[c]{@{}c@{}}Des. house, \\  Foundry\end{tabular} & No                                                         & \begin{tabular}[c]{@{}c@{}}Nearly zero overheads, \\ Non-destructive method\end{tabular}                                                        & \begin{tabular}[c]{@{}c@{}}\rule{0pt}{8pt}Needs reverse engineering for \\ attributes of the Golden chip, \\ Confusion with environmental \\ and process variation effects\end{tabular} \\ \hline

Logic Testing                                                                 &      \cite{21} \cite{25}      & Detection                                                                                                                 & Test                                                           & \begin{tabular}[c]{@{}c@{}}Des. house, \\ Foundry\end{tabular}  & No                                                         & \begin{tabular}[c]{@{}c@{}}Very fast technique, \\ Can be fully automatic\end{tabular}                                                                   & \begin{tabular}[c]{@{}c@{}}\rule{0pt}{8pt}All input combinations are  \\ not covered to activate  HTs,\\ Ineffective when dealing with \\  HTs with sequential triggers\end{tabular}          \\ \hline
\begin{tabular}[c]{@{}l@{}}BISA, \\ Layout Filling\end{tabular}               &       \cite{bisa,40,papa}      & DfHT                                                                                                                       & Back-end                                                       & Foundry                                                        & Yes                                                        & \begin{tabular}[c]{@{}c@{}}\rule{0pt}{8pt}Replacement of filler cells \\  with the functional ones, \\ Independent testing circuit \\ from the original one\end{tabular} & \begin{tabular}[c]{@{}c@{}}Design becomes unroutable \\ in high occupation ratios, \\Favors layouts with \\large continuous gaps\end{tabular}                                                                          \\ \hline
\begin{tabular}[c]{@{}l@{}}Selective \\ Placement\end{tabular}                &       \cite{ASSURER, JohannICCAD, gdsguard}       & DfHT                                                                                                                      & Back-end                                                       & Foundry                                                        & Yes                                                        & \begin{tabular}[c]{@{}c@{}}Lowest overheads among \\ HT prevention techniques\end{tabular}                                               & \begin{tabular}[c]{@{}c@{}}\rule{0pt}{8pt}Risk of violating the critical \\ paths in the routing stage, \\ Timing degradation \end{tabular}                                           \\ \hline
TPAD                                                                          &    \cite{57}        & Both                                                                                                                    & \begin{tabular}[c]{@{}c@{}}Front-end, \\ Back-end\tnote{*} \end{tabular} & \begin{tabular}[c]{@{}c@{}}Des. house, \\ Foundry\end{tabular}  & Yes                                                        & \begin{tabular}[c]{@{}c@{}}\rule{0pt}{8pt}Zero false positives, \\ Applicable to FPGA\end{tabular}                                    & \begin{tabular}[c]{@{}c@{}}\rule{0pt}{8pt}Considerable overheads, \\ Risk of degradation \end{tabular}   \\ \hline

\textbf{SCARF}                                                                         & \textbf{This work} & Both                                                                                                                       & \begin{tabular}[c]{@{}c@{}}Front-end, \\ Back-end\end{tabular} & Foundry                                                        & Yes                                                        & \begin{tabular}[c]{@{}c@{}}\rule{0pt}{8pt}Reusing  verification assets, \\ Low PPA overheads\end{tabular}                                & \begin{tabular}[c]{@{}c@{}}\rule{0pt}{8pt}Achieving high density for \\some designs is challenging\end{tabular}     \\ \hline                                                                                         
\end{tabular}
\begin{tablenotes}
  \item[*]This technique is mainly applied in the front-end stage. To prevent CAD tool attacks, the design is split into two parts, each processed by different CAD tools independently. The final layouts from these tools are later merged in the back-end stage for being set to the foundry for fabrication.
\end{tablenotes}
\end{threeparttable}
\end{table*}

\section{Conclusion}\label{sec:conclusion}

In this paper, we presented a defensive framework aimed at enhancing the security of chips against fabrication-time attacks. The first part of the presented framework focused on the front-end stage, where we demonstrated the reuse of verification assertions by transforming them into security checkers. 
However, some security checkers may not be optimally effective, particularly those designed primarily for debugging purposes. To address this limitation, we proposed a novel methodology involving the insertion of online checkers during the back-end stage of the physical synthesis.

Although we designed the back-end methodology as a complementary approach to the front-end one, both techniques can be used independently based on user preferences and requirements. Our experimental results demonstrate that adding online monitors results in a modest increase of less than 20\% in area, power, and placement density in the worst case. Simultaneously, it enhances the SC by 33.58\% in the best case. Notably, our methodology utilizes only positive slack time during the ECO flow for adding online monitors, ensuring that it does not degrade the overall timing of the design.

While our framework has demonstrated promising results, there is still room for improvement. Future work could focus on embedding various optimizations to reduce the imposed area and power, as well as incorporating path awareness into the flow to make better use of positive slack time.


\bibliographystyle{IEEEtran}
\bibliography{ref}


\begin{IEEEbiography}[{\includegraphics[width=1in,height=1.25in,clip,keepaspectratio]{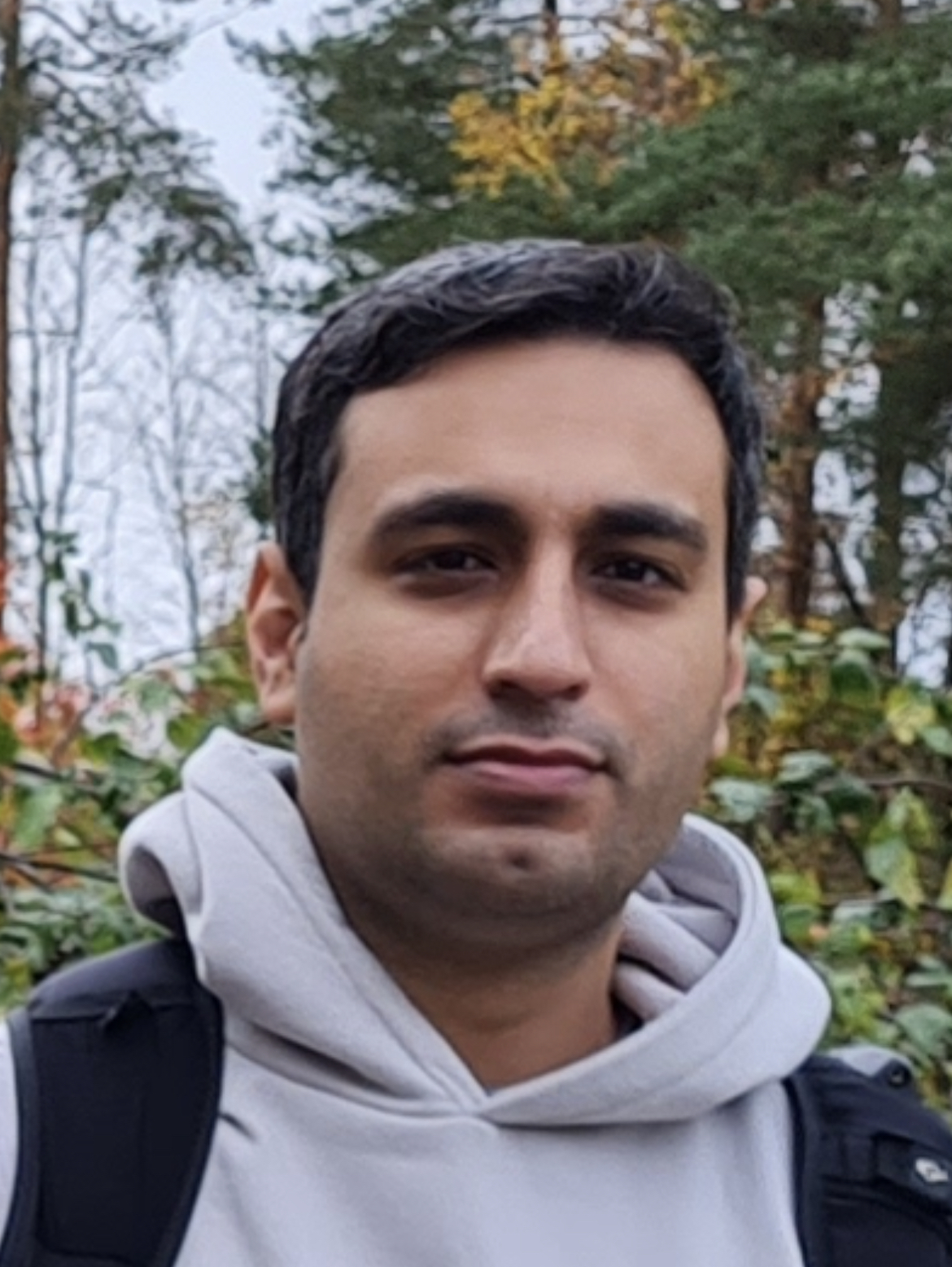}}]{Mohammad Eslami}
received his M.S. degree in computer engineering from the Shahid Bahonar University of Kerman, Kerman, Iran, in 2018. Currently, he is pursuing his doctoral studies at the Centre for Hardware Security, Tallinn University of Technology (TalTech), Tallinn, Estonia. 

His research interests primarily revolve around hardware security, with a particular focus on physical design automation and secure ASIC design.
\end{IEEEbiography}

  \vskip -2\baselineskip plus -1fil

\begin{IEEEbiography}[{\includegraphics[width=1in,height=1.25in,clip,keepaspectratio]{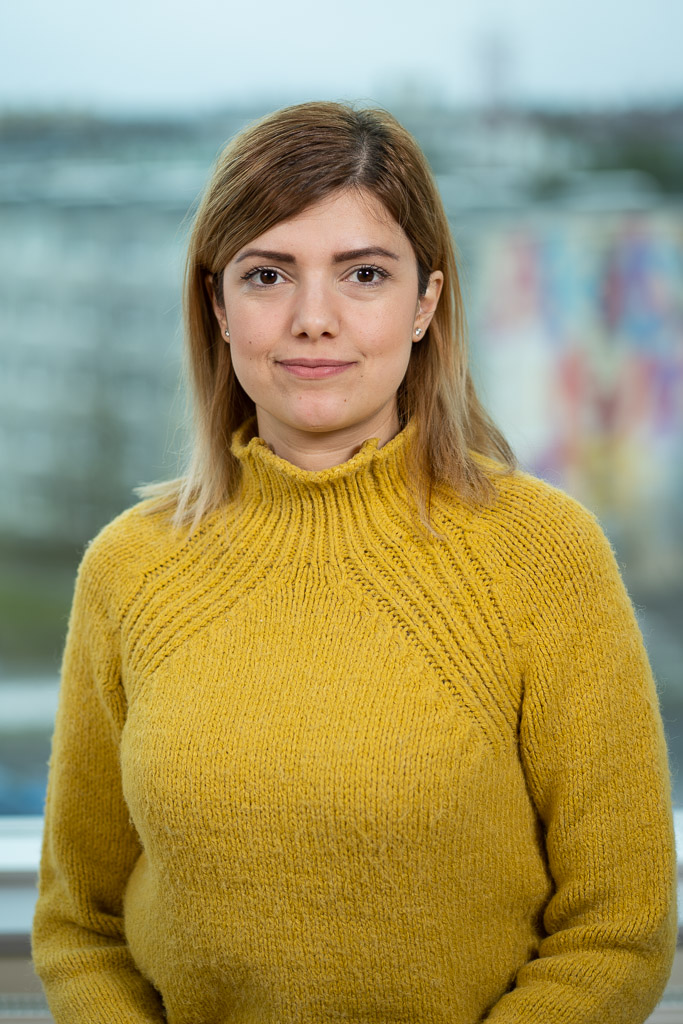}}]{Tara Ghasempouri} received the PhD degree from University of Verona, Italy in 2016.
Currently, she is a Senior Researcher at Tallinn University of Technology (TalTech) in Tallinn, Estonia.
Her fields of interest are Hardware verification and Hardware security. At the moment, she leads projects concerned with security of digital system as well as self-driving vehicle. She started collaboration with Taltech by receiving Mobilitas+ postdoctoral grant. She had research visitor position at the University of Delft, The Netherlands and TU Munich, Germany.
\end{IEEEbiography}

  \vskip -2\baselineskip plus -1fil

\begin{IEEEbiography}[{\includegraphics[width=1in,height=1.25in,clip,keepaspectratio]{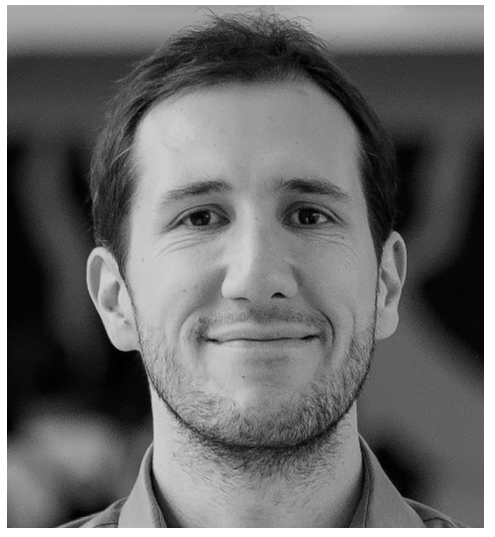}}]{Samuel Pagliarini}
(M'14) received the PhD degree from Telecom ParisTech, Paris, France, in 2013. 

He has held research positions with the University of Bristol, Bristol, UK, and with Carnegie Mellon University, Pittsburgh, PA, USA. From 2019 to 2023, he led the Centre for Hardware Security at Tallinn University of Technology in Tallinn, Estonia. He is currently a professor at Carnegie Mellon University, Pittsburgh, PA, USA.

\end{IEEEbiography}

\vfill

\end{document}